\pdfoutput=1
\documentclass[sigconf]{acmart}
\usepackage[nolist]{acronym}

\usepackage{booktabs}
\usepackage{csquotes}
\usepackage{enumerate}
\usepackage{subcaption}
\usepackage{multirow}
\usepackage{rotating}
\usepackage{pdfpages}
\usepackage{enumitem}
\usepackage{xspace}
\usepackage{tabularx}
\usepackage{float}
\usepackage{pgfplots}
\usepackage{pgfplotstable}
\usepgfplotslibrary{groupplots}
\usetikzlibrary{calc,matrix}
\usepackage{tcolorbox}
\usepackage{enumitem}
\pgfplotsset{compat=1.18}

\usepackage[inline,nomargin,index]{fixme} % Yixin: I copied it from another document, feel free to create your own commands
\fxsetup{status=final,theme=colorsig,mode=multiuser,inlineface=\itshape,envface=\itshape} % hide comments
\FXRegisterAuthor{yz}{ayz}{\color{blue}Yixin}
\renewcommand{\itemautorefname}{Q\kern-0.15em}

\setlist[enumerate,1]{label=(\arabic*), ref=\arabic*}
\setlist[enumerate,2]{label=(\theenumi.\arabic*), ref=\theenumi.\arabic*}
\setlist[enumerate,3]{label=(\theenumii.\arabic*), ref=\theenumii.\arabic*}

\newcounter{qcounter}
\newcommand{\mychoice}[1]{{$\circ$}~#1 \, } % for formatting survey questions in the appendix

\newcolumntype{x}[1]{>{\centering\let\newline\\\arraybackslash\hspace{0pt}}p{#1}}
\newcolumntype{L}[1]{>{\raggedright\let\newline\\\arraybackslash\hspace{0pt}}m{#1}}
\newcolumntype{C}[1]{>{\centering\let\newline\\\arraybackslash\hspace{0pt}}m{#1}}
\newcolumntype{R}[1]{>{\raggedleft\let\newline\\\arraybackslash\hspace{0pt}}m{#1}}

\newcommand{\ie}{i.\,e.}
\newcommand{\eg}{e.\,g.}

\newcommand{\etal}{et~al.\@\xspace}

\newcommand{\infosymbol}{\textcircled{\footnotesize i}\xspace}

\definecolor{c1}{HTML}{648FFF}
\definecolor{c2}{HTML}{785EF0}
\definecolor{c3}{HTML}{DC267F}
\definecolor{c4}{HTML}{FE6100}
\definecolor{c5}{HTML}{FFB000}

\definecolor{LightGreen}{HTML}{e6ecff}
\definecolor{DarkGreen}{HTML}{1e0b75}

\newlist{questions}{enumerate}{1}
\setlist[questions,1]{label={\textbf{RQ\arabic*:}},ref={RQ\arabic*},left=0pt,labelsep=0.5em,listparindent=\parindent}

\usepackage{mdframed}

\usepackage{hyperref}

\AtBeginDocument{%
  }

\setcopyright{acmlicensed}
\copyrightyear{2025}
\acmYear{2025}
\acmDOI{XXXXXXX.XXXXXXX}

\acmISBN{978-1-4503-XXXX-X/18/06}

\renewcommand\footnotetextcopyrightpermission[1]{}
\setcopyright{none}

\begin{document}

\begin{acronym}
    \acro{AI}{artificial intelligence}
    \acro{ALU}{arithmetic logic unit}
    \acro{API}{application programming interface}
    \acro{APU}{application processing unit}
    \acro{ASIC}{application-specific integrated circuit}
    \acro{ATI}{affinity for technology interaction}
    \acro{AXI}{advanced extensible interface}
    \acro{AIC}{Akaike information criterion}

    \acro{BRAM}{block \acs{RAM}}
    \acro{BIC}{Bayesian information criterion}

    \acro{CPU}{central processing unit}
    \acro{CRT}{chinese remainder theorem}
    
    \acro{DL}{deep learning}
    \acro{DLL}{dynamic link library}
    \acroplural{DLL}[DLLs]{dynamic link libraries}
    \acro{DNN}{deep neural network}
    \acro{DPU}{deep learning processing unit}
    \acro{DRM}{digital rights management}
    \acro{DSP}{digital signal processor}
    \acro{DSR}{desired success rate}
    
    \acro{ECC}{elliptic-curve cryptography}
    \acro{ECU}{electronic control unit}
    \acro{EDA}{electronic design automation}
    \acro{E2EE}{end-to-end encryption}

    \acro{FF}{flip-flop}
    \acro{FPGA}{field-programmable gate array}
    \acro{FSM}{finite state machine}
    
    \acro{GCD}{greatest common divisor}
    \acro{GDPR}{General Data Protection Regulation}
    \acro{GPU}{graphics processing unit}

    \acro{HDL}{hardware description language}
    \acro{HCI}{human-computer interaction}
    \acro{HSM}{hardware security module}
    
    \acro{IC}{integrated circuit}
    \acro{IEEE}{institute of electrical and electronics engineers}
    \acro{ICC}{intra-class correlation}
    \acro{IoT}{internet of things}
    \acro{IP}{intellectual property}
    \acro{IRB}{institutional review board}

    \acro{LUT}{lookup table}
    
    \acro{MATE}{man-at-the-end}
    \acro{MITM}{man-in-the-middle}
    \acro{ML}{machine learning}
    \acro{MUX}{multiplexer}

    \acro{OEM}{original equipment manufacturer}

    \acro{PCB}{Printed Circuit Board}
    \acro{PDK}{process design kit}
    \acro{PL}{programmable logic}
    \acro{PKI}{public key infrastructure}
    \acro{PS}{processing system}
    
    \acro{RAM}{random-access memory}
    \acro{ROM}{read-only memory}
    \acro{RQ}{research question}
    \acro{RTL}{register transfer level}

    \acro{SEM}{scanning electron microscope}
    \acro{SGD}{stochastic gradient descent}
    \acro{SoC}{system-on-chip}
    \acroplural{SoC}[SoCs]{systems-on-chip}
    \acro{SSD}{solid-state drive}

    \acro{TEE}{trusted execution environment}

    \acro{VPN}{virtual private network}

    \acro{XAI}{explainable \acl{AI}}
    \acro{XIR}{Xilinx intermediate representation}
\end{acronym}

\title[Exploring End Users' Understanding and Information Needs Regarding Microchips]{\enquote{Make the Voodoo Box Go Bleep Bloop:} Exploring End Users' Understanding and Information Needs Regarding Microchips}


\author{Julian Speith}
\email{julian.speith@mpi-sp.org}
\orcid{0000-0002-8408-8518}

\affiliation{%
  \institution{MPI-SP}
  \city{Bochum}
  \country{Germany}
}

\author{Steffen Becker}
\email{steffen.becker@rub.de}
\orcid{0000-0001-7526-5597}

\affiliation{%
  \institution{Ruhr University Bochum \& MPI-SP}
  \city{Bochum}
  \country{Germany}
}

% \affiliation{%
%   \institution{MPI-SP}
%   \country{Germany}
% }

\author{Timo Speith}
\email{timo.speith@uni-bayreuth.de}
\orcid{0000-0002-6675-154X}

\affiliation{%
  \institution{University of Bayreuth}
  \city{Bayreuth}
  \country{Germany}
}

\author{Markus Weber}
\email{markus.weber@rub.de}
\orcid{0000-0001-7775-807X}

\affiliation{%
  \institution{Ruhr University Bochum}
  \city{Bochum}
  \country{Germany}
}

\author{Yixin Zou}
\email{yixin.zou@mpi-sp.org}
\orcid{0000-0002-9088-705X}

\affiliation{%
  \institution{MPI-SP}
  \city{Bochum}
  \country{Germany}
}

\author{Asia Biega}
\email{asia.biega@mpi-sp.org}
\orcid{0000-0001-8083-0976}

\affiliation{%
  \institution{MPI-SP}
  \city{Bochum}
  \country{Germany}
}

\author{Christof Paar}
\email{christof.paar@mpi-sp.org}
\orcid{0000-0001-8681-2277}

\affiliation{%
  \institution{MPI-SP}
  \city{Bochum}
  \country{Germany}
}

%% Short List of Authors
\renewcommand{\shortauthors}{Speith,~\etal}

\begin{abstract}
Microchips are fundamental components of modern electronic devices, yet they remain opaque to the users who rely on them daily. This opacity, compounded by the complexity of global supply chains and the concealment of proprietary information, raises significant security, trust, and accountability issues. We investigate end users' understanding of microchips, exploring their perceptions of the societal implications and information needs regarding these essential technologies. Through an online survey with 250 participants, we found that while our participants were aware of some microchip applications, they lacked awareness of the broader security, societal, and economic implications. While our participants unanimously desired more information on microchips, their specific information needs were shaped by various factors such as the microchip's application environment and one's affinity for technology interaction. Our findings underscore the necessity for improving end users' awareness and understanding of microchips, and we provide possible directions to pursue this end.
\end{abstract}

% keywords and CCS concepts



\maketitle

\section{Introduction}
\label{xhw_study::sec::introduction}

At the core of the digital revolution are microchips, tiny electronic devices that store and process digital data.
A microchip contains numerous nanometer-sized electronic components (\eg, transistors) on a single piece of semiconductor material (typically silicon). 
These components work together to perform digital processing tasks such as executing computations (\acsp{CPU}, \acsp{GPU}), storing data (\acsp{SSD}, \acs{RAM}), or cryptographic and \acs{AI} acceleration.
Microchips serve as the basic building blocks in the electronic devices we use every day, including smartphones, vehicles, and medical equipment.

As microchips have become ubiquitous and are increasingly being used in critical areas, their geopolitical importance is growing. 
However, due to their rising complexity~\cite{Burrell2016How, DBLP:conf/re/MannCKSS23,apple2022m1ultra}, a globally distributed supply chain~\cite{weste2015cmos,lienig2020fundamentals}, and intentional concealment to protect trade secrets, microchips are often regarded as highly opaque.
This opacity can make it challenging to identify potential safety and security issues, thereby complicating efforts to build trust in these technologies.
Consequently, several concerns regarding microchips have not yet been resolved.
For instance, microchips are susceptible to attacks from a diverse range of adversaries. 
They can be manipulated through hardware Trojans~\cite{Adee2008Hunt}, particularly when employed in safety- and security-critical tasks such as encryption~\cite{DBLP:conf/crypto/KocherJJ99, DBLP:conf/host/DaRoltNFR11}. 
Similarly, previous studies have demonstrated how security issues in the hardware~\cite{Adee2008Hunt,Becker2013Stealthy,Lipp2018Meltdown} can impact the security of end-user devices~\cite{CVE-2023-38606}. 

In response to these concerns, numerous countries have introduced subsidies and regulations to bolster domestic microchip industries~\cite{uschips2022, euchips2022}.
These measures aim to secure production, promote innovation, and foster talent while, at the same time, addressing global supply chain vulnerabilities and security threats.
However, their primary goals are tied to geopolitical strategy and achieving or maintaining technological leadership, underscoring the high stakes in the global microchip race.

Despite the focus on industry and geopolitics, one crucial stakeholder often overlooked in these regulatory discussions is the end user. 
The question arises: should users be considered, and perhaps studied, as integral stakeholders in the microchip ecosystem? 
We think the question is worth exploring because end users are already constantly interacting with and relying on the proper functioning of microchips in their daily lives, albeit often unknowingly and indirectly. 
While end users may be familiar with the fact that the \acsp{CPU} within their computers are microchips, the application of microchips to other technologies and devices might be more hidden. 
Modern cars are built from hundreds of microchips, and smartphones and laptops contain dozens.
Microchips are also increasingly found in medical equipment like insulin pumps, pacemakers, and ventilators---technologies on which someone's life may depend.

We see the potential that improving end-users' understanding of microchips can lead to numerous benefits, such as making more informed product choices, which often start with functionality.
For instance, many people compare CPUs before purchasing a computer.
Some vendors even make microchips the centerpiece of their marketing, such as Apple with its A- and M-series microchips.
However, product choices can also be influenced by factors such as security, trustworthiness, and sustainability.
In this context, the (country of the) manufacturer, materials used, and power consumed during microchip production~\cite{DBLP:conf/hpca/GuptaKLTLW0W21,WILLIAMS20042,VILLARD201598} could become key considerations for product choice~\cite{LIN201211}.

Research in other contexts shows that limited understanding of technologies like the Internet~\cite{kang2015my}, Wi-Fi~\cite{klasnja2009wi}, or home computer security~\cite{wash2010folk} can lead to a false sense of security and inadequate protective practices. 
However, to the best of our knowledge, the academic community has yet to study end-user understanding of, information needs concerning, and trust in microchips.
In light of this gap, we seek to answer the following \acp{RQ}:
\begin{itemize}
    \item \textit{RQ1 [Understanding]} 
    How do end users currently understand microchips?
    \item \textit{RQ2 [Desiderata]} 
    What do end users value concerning and what do they desire to know about microchips?
    \item \textit{RQ3 [Information Needs]}  
    What factors shape end users' information needs when it comes to microchips?
\end{itemize}

To answer these \acp{RQ}, we conducted and evaluated an online survey with 250 end-user participants. 
Our key findings include:
\begin{itemize}
    \item \textbf{End-User Understanding of Microchips.} 
    Participants had a basic understanding of what microchips are and where they are used. 
    However, we also found several misconceptions, and participants mentioned little about the %broader societal impacts of microchips or 
    security and privacy implications of microchips.
    
    \item \textbf{Desirable Properties of Microchips.} 
    When prompted, participants rated cyber security and trustworthiness as their most valued objectives for microchips. 
    At the same time, participants rated safety, accountability, and ethical standards still as \enquote{very important} on average.
    
    \item \textbf{Factors Shaping End Users' Information Needs.} 
    Our participants indicated that they want to know more about microchips and are willing to invest time to that end. 
    The exact type of information they wished for depends on the microchip's specific application environment as well as the participant's affinity for technology interaction.
\end{itemize}

Finally, we discuss interesting patterns from our findings that call for further investigation. 
For example, based on our results, we find that the goals of ongoing political initiatives around microchips might not serve the needs of end users. 
Our study lays the foundation for future research to more thoroughly look into end users' mental models of microchips and design mechanisms that effectively convey information about microchips to end users.
\section{Related Work}
\label{xhw_study::sec::background}

\subsection{User Understanding and Transparency}
Within the usable security and privacy community, past research has studied end users' understanding of \ac{E2EE}~\cite{DBLP:conf/eurousec/SchaewitzLSR21,DBLP:conf/soups/WuZ18}, HTTPS~\cite{DBLP:conf/sp/KrombholzBP0Z19}, home computer security~\cite{wash2010folk}, the Internet~\cite{kang2015my}, online behavioral advertising~\cite{DBLP:conf/cscw/YaoR017}, \acp{VPN}~\cite{DBLP:conf/uss/BinkhorstFKPL22,DBLP:conf/uss/RameshVE23}, and more. 
Some studies further draw the line between non-expert end users and experts such as system administrators and developers~\cite{DBLP:conf/uss/BinkhorstFKPL22,DBLP:conf/sp/KrombholzBP0Z19}. 
Misconceptions are common and often have downstream effects on users' behaviors. 
For example, Renaud~\etal~\cite{DBLP:conf/pet/RenaudVR14} found that incomplete threat models and a general lack of understanding of the email architecture are possible explanations for the low adoption of \ac{E2EE} for emails.
Importantly, there is no perfectly correct understanding~\cite{wash2010folk}, and even experts (with a deeper technical understanding of the technology) can still hold false beliefs~\cite{DBLP:conf/uss/BinkhorstFKPL22, DBLP:conf/sp/KrombholzBP0Z19}.

Studies on end-user understanding contribute insights into theirmisconceptions~\cite{rader2020have,DBLP:conf/soups/WuZ18,DBLP:conf/cscw/YaoR017,kang2015my} and reasoning processes behind threat models~\cite{DBLP:conf/eurousec/SchaewitzLSR21}, which then inform recommendations for how to encourage a secure use of the technology (\eg, through training, better communication, or system design changes)~\cite{DBLP:conf/uss/BinkhorstFKPL22}. 
The mental model approach is often used to describe the model in one's mind about how things work~\cite{wash2010folk}, usually with metaphors from already known domains~\cite{DBLP:journals/ijmms/StaggersN93}. 
For example, Stransky~\etal~\cite{DBLP:conf/soups/StranskyWSHAFWU21} compared six visualizations of security mechanisms for messaging apps based on users' mental models of \ac{E2EE}, finding that simple text disclosures were sufficient, yet user perceptions were more fundamentally shaped by preconceived expectations. 
Other work has sought to build visualization dashboards~\cite{DBLP:journals/popets/ReitingerWMU24,DBLP:journals/popets/FarkeBGA24} and design probes~\cite{DBLP:journals/imwut/BarbosaWUW21} to improve users' understanding of online tracking and inferences. 
Researchers have also explored using labels to convey the data practices of \ac{IoT} devices~\cite{DBLP:journals/ieeesp/NaeiniDAC22} and mobile apps~\cite{DBLP:conf/soups/ZhangKN0C24} to help consumers make purchase decisions, and such initiatives have received buy-ins from industry players and regulators~\cite{DBLP:journals/cacm/CranorAN24}.

Parallel efforts exist in the \acs{XAI} community, where the focus is to unpack the black box of \acs{AI}-based systems to end users, making the decision-making more understandable and transparent~\cite{DBLP:conf/fat/Speith22}. 
An individual's understanding of an \acs{AI}-based system can be increased by \enquote{white-box} explanations (\ie, that show the inner workings of an algorithm)~\cite{DBLP:conf/chi/ChengWZOGHZ19}, contextualizing general terminologies~\cite{DBLP:journals/pacmhci/ShenJCPZH20}, showing each feature's contribution to the model's prediction~\cite{DBLP:conf/iui/WangY21}, among other techniques.
The understanding can also be affected by the individual's domain expertise in the decision-making task~\cite{DBLP:conf/iui/WangY21} as well as the explanation's modality (\eg, textual, visual, or interactive)~\cite{DBLP:conf/fat/SchmudeKMT23}. 
Speith~\etal~\cite{DBLP:conf/re/SpeithSBZBP24} connect explainability to hardware in the context of requirements engineering, with a particular focus on microchips.
Among their future research directions, they explicitly propose to explore end-users' mental models of microchips.

Against these backgrounds, we see the potential that a better understanding of microchips can benefit end users. Our study provides novel knowledge of end users' current understanding of microchips and their informational wants, laying the foundation for future work on transparency mechanisms and educational efforts.

\subsection{Studies on Microchip (Security)}
To the best of our knowledge, there has been no prior work on end-user understanding of and interactions with microchips. That being said, prior research has examined the relationship between users and various microchip-based technologies, including autonomous vehicles~\cite{DBLP:conf/chi/TranPHWT24,DBLP:conf/chi/ChangCDCYCZZG24,DBLP:conf/chi/ChuZSGLGDZ23}, drones~\cite{DBLP:conf/chi/DongZCCL24}, robots~\cite{DBLP:conf/chi/SchneidersBCMCN24,DBLP:conf/chi/LuoDK24}, smart home devices~\cite{DBLP:conf/chi/ChiangKBC24}, and sensors in smart cities~\cite{DBLP:conf/chi/CorbettD24,DBLP:conf/chi/WindlW0M23}. 
These studies collectively contribute to our understanding of how users interact with and perceive emerging microchip technologies.

Research has also focused on improving the design and sustainability of \acfp{PCB}.\footnote{A \ac{PCB} is a flat surface that electrically connects electronic devices such as microchips.} 
Lin~\etal~\cite{DBLP:conf/chi/LinRPTDHM24} highlighted design space exploration as a promising alternative to fully automated or manual \ac{PCB} design approaches.
Yan~\etal~\cite{DBLP:conf/chi/YanL0P24} proposed SolderlessPCB to enhance the reusability of electronic components by eliminating the need for soldering components onto the \ac{PCB}.
Similarly, Arroyos \etal~\cite{DBLP:conf/chi/ArroyosVKOSSIN22} presented a functional computer mouse made from biodegradable \ac{PCB} materials, demonstrating that these components can dissolve in water, which allows for the reuse of mounted microchips.
Strasnick~\etal~\cite{DBLP:conf/chi/StrasnickAF21} introduced a \ac{PCB} debugging tool that aids in analog circuit debugging by facilitating the comparison between the physical circuit and a simulated model.

Focusing on security research, a few usable security papers have touched upon the role of hardware, although the findings were often discussed in passing as a small part of the main insights.
For example, Schmüser~\etal~\cite{DBLP:conf/chi/SchmuserRWSB0SW24} conducted a study on online security advice during the Ukraine war and found that the Twitter community regarded hardware as a medium-level concern, which was discussed primarily in the context of locking devices, disabling biometrics, and turning off location services.
Similarly, Gallardo~\etal~\cite{DBLP:conf/chi/GallardoEBBC24} discovered that security experts and energy system operators tend to underestimate the risks associated with hardware-based attacks.
Yu~\etal~\cite{DBLP:conf/chi/YuSDW24} found that while cryptocurrency users prefer hardware wallets for security reasons, they often refrain from using them due to usability challenges.
Reynolds \etal~\cite{DBLP:conf/sp/ReynoldsSRDRS18} highlighted usabilty issues in setting up YubiKeys (\ie hardware security tokens for two-factor authentication) with Google, Facebook, and Windows accounts.
Pfeffer \etal~\cite{DBLP:conf/uss/PfefferMDGSWFK21} later surveyed the effectiveness and usability of authenticity checks for such tokens, finding that users often neglect these essential checks, thereby undermining the security guarantees of the tokens.

Previous research has also explored the role of users in the security assurance of microchips~\cite{DBLP:conf/soups/0001WARP20,DBLP:journals/tochi/WiesenBWPR23,DBLP:conf/chi/WalendyWL0WE0FR24}. 
These studies examine the cognitive processes~\cite{DBLP:conf/soups/0001WARP20} and strategies~\cite{DBLP:journals/tochi/WiesenBWPR23} involved in hardware reverse engineering, employing methods such as eye tracking and think-aloud protocols~\cite{DBLP:conf/chi/WalendyWL0WE0FR24} to gain insights on how users interact with and analyze microchips.

While these works offer valuable insights into user interactions with hardware, our study goes beyond the technical aspects of hardware and broadens this inquiry by focusing on end-users' understanding of microchips, perceptions of their broader societal and security implications, and end-user information needs.
\section{Methods}
\label{xhw_study::sec::methods}
We conducted an online survey with 250 participants recruited via Prolific. 
A core part of the survey is a vignette setup: to make the concept of microchips less abstract and more accessible, we presented the participants with five scenarios based on real-world applications of microchips.
Each vignette consisted of a setting that describes a particular use case of a microchip and a desideratum (\ie, a property that might be desirable to end users).
Following the vignette description, we asked participants to rate the importance of the desideratum as well as the importance of receiving specific types of information about the microchip in the respective scenario.
Below, we outline our rationale for selecting vignette components and information types, present details of questionnaire design and study procedures, and address ethical considerations and data analysis techniques.

\subsection{Topic Selection and Item Generation}
\label{xhw_study::subsec::topic}
As the first step of scoping the survey, we identified five concrete settings in which microchips may be used, five desiderata that end users may want satisfied for microchips, and five kinds of information presented to end users.
The settings, desiderata, and information types were derived
from a literature review and discussions among experts, and refined in pilot studies (see \autoref{xhw_study::subsec::implementation}).

\paragraph{Derivation of settings involving microchips}
To help end users relate to microchips, we selected five settings in which microchips are employed.
We deliberately selected settings across a diverse range of applications, touching on aspects that end users may encounter in their everyday lives.
Specifically, we consider microchips (i)~\textit{controlling the entertainment system in a car}, (ii)~\textit{enabling wireless communication in a cell tower}, (iii)~\textit{controlling a pacemaker to maintain an adequate heart rate}, (iv)~\textit{enabling fingerprint unlocking of a smartphone}, and (v)~\textit{controlling the steering of an airplane}.

\paragraph{Derivation of end-user desiderata}
In our survey, we consider different goals that are desirable for end users.
We borrow an initial set of desiderata from literature on other technical systems~\cite{Chazette2021Exploring,Langer2021What,speith2023building}.
Through pilot testing (see \autoref{xhw_study::subsec::implementation}), we narrowed down the selection to five desiderata that are relevant for microchips and at the same time relatable to end users:
(a) \textit{accountability}, (b)~\textit{safety}, (c)~\textit{cyber security}, (d)~\textit{trustworthiness}, and (e)~\textit{ethical standards}.

\paragraph{Derivation of information facilitating microchip understanding}
\label{xhw_study::par::information}
The five different kinds of information offered to the end user are derived from different stages of the microchip design and manufacturing process~\cite{weste2015cmos,lienig2020fundamentals}.
Microchips are designed using a high-level language similar to regular programming languages.
The design descriptions are then implemented as an electronic circuit using automated software tools.
Next, the  design is handed to the manufacturer who produces the microchip in one of their production facilities, also known as \emph{fabs}~\cite{geiger1990vlsi}.
Derived from this process, we list the following as information to provide about microchips: (1)~\textit{who designed and manufactured the microchips} and (2)~\textit{how the microchips were designed and manufactured}.

Especially safety- and security-critical microchips must be certified by independent government bodies or dedicated testing service providers before use. 
As such, another useful piece of information could be (3)~\textit{how the microchips have been approved for use}.

The fabricated chip is finally integrated into a device such as a smartphone, a pacemaker, or a car.
Therefore, further relevant information could be (4)~\textit{how the microchips interact with the system} and (5)~\textit{which functionality the microchips provide}.

\subsection{Questionnaire Design}
\label{xhw_study::subsec::questionnaire}
We drew from our team members' expertise in usable security and embedded systems when designing the questionnaire. 
We took care to make our questionnaire understandable to end users through several rounds of piloting.
In the following, we briefly describe the flow of our questionnaire (see \autoref{xhw_study::app::survey} for a full version).

\subsubsection{Introduction}
At the beginning, we stated the purpose of our study, the expected duration of 25 minutes, and provided information on data handling and data protection.
Before participants could proceed, we asked them to give informed consent and to confirm that they were residents of the United States and at least 18 years of age~(\autoref{xhw_study::question::consent}).
Next, we asked participants nine questions on a six-point Likert scale to assess their tendency to actively engage in technology interaction using a validated psychometric scale~(\autoref{xhw_study::question::ati})~\cite{franke2019personal}.

\subsubsection{General questions on microchips}
In an open question, we asked participants what comes to mind when thinking of microchips (\autoref{xhw_study::question::perception}).
Further, we asked them whether or not they would like to understand more about microchips and invited them to give reasons for their choice~(\autoref{xhw_study::question::understand_more}).
We also queried participants regarding the time they would be willing to invest to better understand microchips~(\autoref{xhw_study::question::time_willing_before}) and on the time they currently invest for the same purpose~(\autoref{xhw_study::question::time_actual}), both on a five-point scale.
Next, we provided some background on microchips to align participants' basic understanding~(\autoref{xhw_study::question::background}).
To conclude this block, we presented five settings in randomized order involving microchips and asked participants to rate their criticality as the impact that a microchip malfunction would have on the participant themselves~(\autoref{xhw_study::question::criticality}).

\subsubsection{Vignettes}
From the 25 possible combinations of settings and desiderata (see \autoref{xhw_study::subsec::topic}), we formed five sets of five vignettes each, in which each setting and each desideratum occurs only exactly once.
At the core of our questionnaire, we showed participants one of these sets.
An example vignette is shown in \autoref{xhw_study::question::v0}.
For each vignette, we first asked participants to rate the importance of having a high level of the respective desideratum in the setting at hand on a five-point Likert scale~(\autoref{xhw_study::question::v0_desideratum}).
We then invited participants to explain their choices in an open-ended response~(\autoref{xhw_study::question::v0_desideratum_open}).
Second, we asked participants to rate the importance of receiving each of the five types of information (see \autoref{xhw_study::subsec::topic}) to assess the given desideratum in the specified setting on a five-point Likert scale~(\autoref{xhw_study::question::v0_information}).
Subsequently, we requested them to briefly explain their choice for one of the information types in an open-ended response~(\autoref{xhw_study::question::v0_information_open}).
Throughout the vignettes, we provided tooltips for some phrases (see \textbf{\textcolor[HTML]{B6321C}{red}} parts in \autoref{xhw_study::question::v0}) that, once hovered over with the cursor, would explain desiderata and types of information in simple language so that participants' mental models of these items are aligned to our understanding and they could get clarifications as needed as they completed the survey.

\subsubsection{Comprehension check}
To determine whether participants actually understood the desiderata, we presented them with an assignment exercise that asked them to match five randomly ordered sentences indicating the meaning of a desideratum to the desideratum in question~(\autoref{xhw_study::question::properties}).
The content of the sentences was based on the tooltips for the desiderata from the vignettes.
We again asked participants about their willingness to invest time in understanding microchips~(\autoref{xhw_study::question::time_willing_after})
to see whether it has changed compared to before~(\autoref{xhw_study::question::time_willing_before}).

\subsubsection{Demographics}
We asked for participants to indicate their gender~(\autoref{xhw_study::question::gender}), age range~(\autoref{xhw_study::question::age}), highest level of education~(\autoref{xhw_study::question::education}), and whether they had any prior practical experience with microchips~(\autoref{xhw_study::question::experience}).
Finally, we inquired if our participants had any feedback or anything they would like to share with us~(\autoref{xhw_study::question::feedback}).

\subsection{Survey Implementation}
\label{xhw_study::subsec::implementation}
We implemented our questionnaire using Qualtrics and recruited US-based English-speaking participants via Prolific.

\subsubsection{Pilot Testing} 
We conducted several pilot studies with a total of 79 participants to ensure end-user comprehension---specifically of the desiderata---by analyzing the open-ended questions~\autoref{xhw_study::question::v0_desideratum_open} and \autoref{xhw_study::question::v0_information_open} on participants' assessment of~\autoref{xhw_study::question::v0_desideratum} and \autoref{xhw_study::question::v0_information} with respect to misunderstandings.
Through these pilots, we aimed to determine whether our desiderata are indeed relevant to and comprehensible for end users, and if we had missed any desiderata that were important to them.
The extensive piloting led to several iterations of the questionnaire, particularly in terms of wording, sharpening of information types, and exclusion of unclear or irrelevant desiderata.

\subsubsection{Data Collection} 
We rolled out the main study with a gender-balanced sample of 250 participants over 10 days by releasing slots to batches of 25 participants, each at different times of the day.
The sample size of 250 was determined using a power analysis for multiple regression models.
We aimed for the detection of a small effect size $f^2$$=$$0.15$, power$=$$0.95$, and a significance level of $\alpha$$=$$0.05$.
For our power analysis, we indicated a total of 27 predictors, which is the sum of the number of vignettes, one's \ac{ATI} score~(see \autoref{xhw_study::question::ati})~\cite{franke2019personal} and whether or not participants would like to understand more about microchips (see \autoref{xhw_study::question::understand_more}).
%As we qualitatively analyzed participants' open-ended responses, we identified and excluded five submissions where the responses were obviously generated by \acs{AI}.
Participants took a median time of 24:19 minutes to complete our questionnaire and were compensated with 7.50~GBP, thus an hourly wage of 18.51~GBP.

\subsection{Ethics and Data Protection}
\label{xhw_study::subsec::ethics}
We could not have our planned study fully reviewed by an ethics committee because our department did not operate an \ac{IRB} at the time.
However, we reviewed our study in line with the application form for ethical approval of human studies from another department and reached the conclusion that our study would be \ac{IRB}-exempt in their case.
In addition, by limiting the survey to a few demographic questions, notably not asking about region of residence, we ensured the anonymity of our participants from the beginning. 
All data collected were stored on our institution's own servers, to which only the researchers involved in the project have access.

\subsection{Data Analysis}
\label{xhw_study::subsec::analysis}

\subsubsection{Qualitative Analysis}
To obtain insights into end-user perceptions of microchips~(\autoref{xhw_study::question::perception}) and their willingness to understand more about microchips~(\autoref{xhw_study::question::understand_more}), we conducted qualitative analysis~\cite{mayring_qualitative_2014} of the open-ended responses.
To this end, we used inductive thematic analysis.
The coding was executed by two coders, one with a background in hardware security and the other in computer science and \acs{AI} ethics.
Both coders first independently coded 50 responses (20\%). 
Each response could be assigned one or more codes.
Both coders then discussed their results and agreed on a common codebook for each of the two open questions.
In the process, the codebooks were refined through discussions among the coders by deleting, merging, and adding codes.
In the end, the final codebook for \autoref{xhw_study::question::perception} contained 57 codes while the one for \autoref{xhw_study::question::understand_more} comprised 24 codes.
They then both applied these codebooks to the remaining 200 responses (80\%).
To measure inter-coder reliability, Krippendorff's alpha~\cite{krippendorff2018content} was computed over all codes based on the MASI distance~\cite{passonneau2006measuring} between codes assigned by both coders.
This resulted in $\alpha$$=$$0.71$ for \autoref{xhw_study::question::perception} and $\alpha$$=$$0.76$ for \autoref{xhw_study::question::understand_more}, indicating \textit{substantial agreement} between the coders~\cite{landis1977measurement}.
Finally, both coders discussed discrepancies in their code assignments and fully agreed on a common coding.

\subsubsection{Statistical Analysis}
We applied descriptive statistics to describe the sample and overall trends regarding participants' perception of the importance of different scenarios, desiderata, as well as their affinity for technology interaction~\cite{franke2019personal}.
To explain participants' perceived importance of desiderata in different scenarios and information that might facilitate microchip understanding, we utilized inferential statistics.

For the perceived importance of desiderata in different scenarios, we used multiple linear regression models with dummy variables.
It is reasonable to assume that the ratings given by an individual participant are more similar than those given between participants. 
Therefore, we used multilevel modeling for this analysis.
We calculated \ac{ICC}~\cite{Koch2006}, or in this case, intra-individual correlation with intercept-only models. 
As \ac{ICC} accounts for $36\% - 45\%$ of the overall variance, we decided to use random-intercept models for further linear regression analysis.
For the random-intercept models, we calculated \textit{marginal} $R^2$ as well as \textit{conditional} $R^2$~\cite{Nakagawa2012}.
Marginal $R^2$ considers only the variance of the fixed effects, while the conditional $R^2$ takes both the fixed and random effects---in this case, participant ID---into account.
By subtracting marginal $R^2$ from conditional $R^2$, the contribution of the random effects can be calculated.
For model comparisons, we have also considered \ac{AIC}, \ac{BIC}, and deviance.

For all regression models, we applied Bonferroni corrections to take into account the probability of observing a false positive (\ie, a type I error).
In other words, we considered regression coefficients statistically significant only when $p$$<$$.001$ ($p$$=$$\alpha/m$ for the Bonferroni correction where $m$ is the number of comparisons, and $m$$=$$55$ when we had five regression models with 11 predictors each). 

\subsection{Limitations}
\label{xhw_study::subsec::limitations}
To the best of our knowledge, we are the first to explore end-user perspectives on microchips.
Accordingly, we had to develop our questionnaire from scratch.
To make the topic more accessible to end users, we decided to present our participants with vignettes.
However, despite careful selection for diversity, our vignette settings can only represent a small sample of the actual applications of microchips.
We also had to make a pre-selection for the desiderata and the types of information we investigated. 
We mitigated the self-selection bias by iteratively checking for missing items from participants' open-ended responses during pilot testing.

It is possible that comprehension issues may arise from the desiderata we provided (especially for similar ones like security and safety): participants may not clearly differentiate between the desiderata, or their understanding of the desiderata might differ from our definitions. 
We included tooltips as well as comprehension checks to address this issue, and our results show that participants correctly matched the descriptions to the respective desideratum in 90\% of all cases.
Participants had more issues comprehending trustworthiness (84\%) than ethical standards (94\%). 
For the other desiderata, comprehension is between 88\% and 92\%.

Further, we conducted our study only with residents of the United States, and our results may not be generalizable to other countries or societies where there may be specific sociocultural and political factors that shape discussions about microchips.
Last, in our survey, we only collected self-reported data about participants' willingness and time spent learning about microchips, which might not accurately reflect their actual behaviors.

\section{Results}
\label{xhw_study::sec::results}

\subsection{Sample Description}
\label{xhw_study::subsec::sample}
While participants' gender distribution is balanced, nearly 80\% of participants were between 18 and 44 years old and about 60\% had a post-secondary education.
The vast majority (92.4\%) of participants indicated that they had no practical experience designing, manufacturing, testing, or deploying hardware and were not involved with the subject at a policy level.
Participants exhibited a high degree of affinity for technology interaction ($M$$=$$4.05$, $SD$$=$$0.91$ on a 5-point scale).
\autoref{xhw_study::tab::demographics} in \autoref{xhw_study::app::demographics} provides detailed demographic information about the 250 participants.

\subsection{RQ1: End User Understanding of Microchips}
\label{xhw_study::subsec::coding_results}

\subsubsection{End-User Perception of Microchips}
\label{xhw_study::subsubsec::perceptions}
We coded participants' responses regarding their perceptions of microchips~(\autoref{xhw_study::question::perception}) as described in \autoref{xhw_study::subsec::analysis} and present the main results below.
See \autoref{xhw_study::tab::codebook_perception} in \autoref{xhw_study::app::codebook} for an overview of all assigned codes.

\paragraph{Participants' Perceptions Center Around Device Types} 
Participants primarily associate microchips with the applications they are deployed in. 
In 104 (42\%) cases, participants mentioned microchips' deployment in computers, followed by phones (47; 19\%), vehicles (22; 9\%), and tablets (8; 3\%). 
In addition to computers themselves, participants occasionally mentioned microchips' functioning as internal computer components (12; 5\%) or even more precisely, CPU (28; 11\%), motherboard (17; 7\%), and memory (13; 5\%). 
Participants also mentioned other devices or systems in 28 (11\%) cases such as robotics, credit cards, and household devices. 

Additionally, 71 (28\%) participants mentioned the broad notion that microchips are widespread and used across devices, using phrases such as \textit{\enquote{They are used in everything}} and \textit{\enquote{They power many things.}} 
Participants also associated microchips with technology (54; 22\%), electronics (50; 20\%), and technological advancement (47; 19\%), using phrases such as \textit{\enquote{they advance in technology constantly.}}

\paragraph{Microchip Shape and Composition}
Apart from the use cases of microchips, 84 (34\%) participants mentioned small size as a property of microchips (\eg, \textit{\enquote{Microchips are incredibly small}}). 
Another 35 (14\%) participants commented on the composition of microchips (\eg, \textit{\enquote{set of electronic circuits on a small piece}} and \textit{\enquote{silicon chips with thousands of [...] transistors}}). 
Furthermore, 22 (9\%) participants referred to the processing power of microchips (\eg, \textit{\enquote{powerhouse of the computer}} and \textit{\enquote{powerful processing system}}). 

\paragraph{Perceived Microchip Functionalities}
In 83 (33\%) cases, participants commented on microchips' general functionality as building blocks that make things work (\eg, \textit{\enquote{main components of personal computers}} and \textit{\enquote{make electronic devices work}}).
Some other participants delved into specific aspects of functionalities such as data storage (35; 14\%), data processing (25; 10\%), and communication capabilities (11; 4\%). 
Another 12 (5\%) participants described microchips as things that enact control (\eg, \textit{\enquote{dictate and command certain functions}}), and 15 (6\%) participants recognized microchips' diverse functionalities (\eg, \textit{\enquote{perform a variety of functions}}).

\paragraph{Misconceptions About Implanting Microchips}
A recurring theme among participants' responses was microchips being implanted into humans (29; 12\%) and animals (27; 11\%), conveyed in phrases such as \textit{\enquote{microchips being put into people}} and \textit{\enquote{inserted into dogs.}} 
This understanding likely comes from \enquote{microchipping} being a common term for animal implants in the United States.
Especially in the context of pets, tracking capabilities of microchips are mentioned in 18 (7\%) cases for \textit{\enquote{locating lost pets.}}
Microchips implanted into humans also co-occurred with conspiracy theories in 11 (4\%) cases.
In particular, six (2\%) participants mentioned microchips in the context of the COVID-19 pandemic (\eg, \textit{\enquote{We were all injected with one with the coronavirus vaccine}}).

\paragraph{Broader Societal and Security Implications Rarely Mentioned}
While microchips are featured prominently in geopolitical debates, supply chain issues about microchips were mentioned only occasionally in 18 (7\%) cases (\eg, \textit{\enquote{They are scarce in many places}} and \textit{\enquote{caused a massive shortage of vehicles}}). 
In particular, foreign manufacturing was identified as an issue in nine (4\%) cases (\eg, \textit{\enquote{Most that we need in America are made in Taiwan}}).
Another 20 (8\%) participants commented on microchips' societal impacts (\eg, \textit{\enquote{they are a major part of society}}) and political aspects (\eg, \textit{\enquote{they passed the CHIPS Act}}).
Only nine (4\%) participants mentioned security and privacy issues related to microchips proactively, commenting that microchips are \textit{\enquote{vulnerable to cybersecurity attacks}} and expressing \textit{\enquote{I have privacy concerns with them.}}

\subsubsection{End User Willingness to Understand More About Microchips}
\label{xhw_study::subsubsec::understand_more}
In \autoref{xhw_study::question::understand_more}, we asked participants whether they wanted to understand more about microchips and to provide reasons for their choice; we further asked participants about their aspirational and current time spent to understand microchips (\autoref{xhw_study::question::time_willing_before}) and ~(\autoref{xhw_study::question::time_actual}). \autoref{xhw_study::tab::codebook_understand} in \autoref{xhw_study::app::codebook} includes an overview of all assigned codes for participants' reasoning, and below we summarize key findings.

\paragraph{Participants Willing to Learn More About Microchips} 
In response to \autoref{xhw_study::question::understand_more}, 76\% of participants stated that they would like to know more about microchips, and 24\% did not want to know more.
When comparing responses to \autoref{xhw_study::question::time_willing_before} and \autoref{xhw_study::question::time_actual}, we observe the trend that participants would like to spend more time understanding microchips (\eg, for a newly acquired device) compared to the time they spend at the moment, reflecting a strong aspiration for learning more about microchips.
We asked participants twice about the time they are willing to spend on better understanding microchips---once at the beginning (\autoref{xhw_study::question::time_willing_before}) and once towards the end of the survey (\autoref{xhw_study::question::time_willing_after})---to check on potential social desirability bias, and we did not see noticeable changes in the responses to these two questions.

\paragraph{Motivation: Gaining Knowledge, as Existing Knowledge is Lacking}
In 96 (38\%) cases, participants mentioned that they want to know more about microchips to gain knowledge in general (\eg, \textit{\enquote{I like learning in general}} and \textit{\enquote{I can expand my knowledge}}). 
Another 20 (8\%) participants stated their motivation came from a desire of wanting to keep up with progress (\eg, \textit{\enquote{to stay up to date on technology}}), and 24 (10\%) expressed interest in following along the scientific progress (\eg, \textit{\enquote{I would love to know how it develops}}).
In addition, 46 (18\%) participants wanted to better understand the functionality of microchips (\eg, \textit{\enquote{I would like to know how they work}}), and 10 (4\%) wanted to better understand the manufacturing processes. 
The motivation to learn more is also related to participants' self-reported lack of existing knowledge, as 24 (10\%) participants stated that they had incomplete knowledge of microchips so far (\eg,\textit{\enquote{they feel a bit like magic}} and \textit{\enquote{I don't know much about them}}).

\paragraph{Motivation: Importance and Influence on (Future) Life}
In 32 (13\%) cases, participants acknowledged that microchips are omnipresent in daily life (\eg, \textit{\enquote{they became more integrated into our everyday lives}}). 
Additionally, 24 (10\%) participants mentioned microchips' impact on society (\eg, \textit{\enquote{what dangers it could bring to society}}) or their importance for the future (\eg, \textit{\enquote{it is a huge part of the future}}). 
Another 16 (6\%) participants commented that they wanted to broaden their understanding because of microchips' inherent link to technology (\eg, \textit{\enquote{I could learn how to better use tech}}). 

\paragraph{Hurdle: Lack of Interest and Need}
Among participants who did not want to know more about microchips, 28 (11\%) expressed that they have no interest in the topic (\eg, \textit{\enquote{I don't care}} and \textit{\enquote{It is a boring topic}}). 
Another 16 (6\%) participants did not see the need to understand more (\eg, \textit{\enquote{I know as much as I need to know about them}} and \textit{\enquote{It's not something I have to deal with a lot}}). 

\paragraph{Hurdle: Satisfaction, Complexity, and Fear}
In 25 (10\%) cases, participants mentioned that they were satisfied with their current level of knowledge about microchips or they would be satisfied as long as the microchips work as intended even if they do not know why (\eg, \textit{\enquote{as long as microchips work I don't care why or how}}). 
A total of 15 (6\%) participants felt that the topic was too complicated (\eg, \textit{\enquote{it sounds too intricate}} and \textit{\enquote{It's too complicated and will hurt my brain}}). 
Another eight (3\%) participants expressed fear regarding microchips in general (\eg,  \textit{\enquote{I worry what will be developed in the future}} and \textit{\enquote{I am afraid of them}}).

\begin{tcolorbox}
\subsubsection*{Summary in Light of RQ1}
\label{xhw_study::subsubsec::revisiting_rq1}
A majority of participants have a basic understanding of what microchips are, where they are deployed, and what they are capable of.
Furthermore, about three-quarters of our participants expressed a desire to learn more about microchips, mostly to expand their knowledge and keep up with the rapid technological advances.
Nevertheless, participants rarely commented on the societal implications of microchips or expressed concerns about the security and privacy aspects.
Thus, we observe that end users have the baseline knowledge and motivation to be involved as stakeholders in the hardware ecosystem, but educational efforts are needed to deepen their existing understanding and address misconceptions.
\end{tcolorbox}

\subsection{RQ2: Importance of Desiderata and Information Types}
\label{xhw_study::subsec::descriptive}

\subsubsection{Criticality of Microchip Application Settings}
\label{xhw_study::subsubsec::setting_criticality}
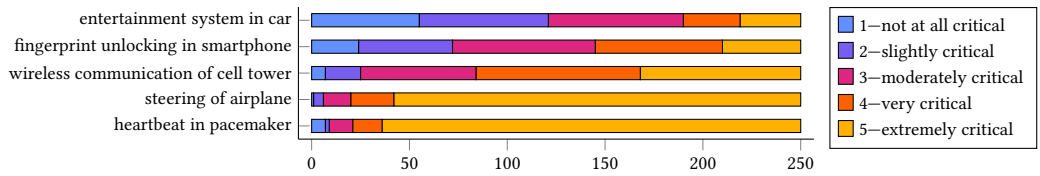
\begin{figure*}[htbp]
    \footnotesize
    \centering
    \begin{tikzpicture}
        \begin{axis}[
            xbar stacked, 
            xmin=0,
            y=10pt,
            bar width=5pt,
            ytick=data, 
            yticklabels from table={\critdata}{Label},
            enlarge y limits={abs=5pt},
            enlarge x limits={abs=5pt},
            axis y line*=left,
            axis x line*=bottom,
            xtick align=outside,
            ytick align=inside,
            legend columns=1,
            legend pos=outer north east,
            legend cell align=left,
        ]
            \addplot [fill=c1] table [x=t1, meta=Label,y expr=\coordindex] {\critdata};
            \addplot [fill=c2] table [x=t2, meta=Label,y expr=\coordindex] {\critdata};
            \addplot [fill=c3] table [x=t3, meta=Label,y expr=\coordindex] {\critdata};
            \addplot [fill=c4] table [x=t4, meta=Label,y expr=\coordindex] {\critdata};
            \addplot [fill=c5] table [x=t5, meta=Label,y expr=\coordindex] {\critdata};
            \legend{1---not at all critical,2---slightly critical,3---moderately critical,4---very critical,5---extremely critical}
        \end{axis}
    \end{tikzpicture}
    \caption{Participants' criticality ratings of the five different settings presented in our survey vignettes.}
    \label{xhw_study::fig::criticality_settings}
    \Description{The figure depicts a horizontal stacked bar chart comprising five bars of five sections each and a legend to the right of the bar chart. On the x-axis, all bars span from 0 to 250. On the y-axis, the first bar (the top one) is labeled "entertainment system in car", the second bar is labeled "fingerprint unlocking in smartphone", the third bar is labeled "wireless communication of cell tower", the fourth bar is labeled "steering of airplane", the fifth (and last) bar is labeled "heartbeat in pacemaker". The legend on the right then assigns meanings to the segments of each bar. It comprises five labels and is organized in two columns of three rows that are to be read from top to bottom first, left to right second. The first label says "1-not at all critical", the second "2-slightly critical", the third "3-moderately critical", the fourth "4-very critical", and the fifth "5-extremely critical".}
\end{figure*}

\autoref{xhw_study::fig::criticality_settings} depicts our participants' perceived criticality of the five settings on a scale from \textit{1---not at all critical} to \textit{5---extremely critical}.
In line with our expectations, the most critical settings were microchips deployed in an airplane ($M$$=$$4.72$, $SD$$=$$0.69$) and in a pacemaker ($M$$=$$4.71$, $SD$$=$$0.83$).
The two settings at the intermediate level were microchips that enable wireless communication in a cell tower ($M$$=$$3.86$, $SD$$=$$1.04$) and microchips that enable fingerprint unlocking in a smartphone ($M$$=$$3.20$, $SD$$=$$1.20$).
Microchips in the entertainment system of a car were rated the least critical ($M$$=$$2.66$, $SD$$=$$1.28$). 

\pgfplotstableread{
Label       t1 t2 t3 t4 t5
heartbeat~in~pacemaker   7 2 12 15 214
steering~of~airplane    1 5 14 22 208
wireless~communication~of~cell~tower  7 18 59 84 82
fingerprint~unlocking~in~smartphone  24 48 73 65 40
entertainment~system~in~car         55 66 69 29 31
}\critdata

A Welch's ANOVA revealed significant differences in the perceived criticality across scenarios ($F(4)$$=$$198.35$, $p$$<$$.001$).
Subsequent pairwise Wilcoxon rank sum tests with Bonferroni corrections revealed significant differences between all scenarios except between the airplane and pacemaker scenarios.

\begin{figure}[htb]
    \centering
    \footnotesize
    \begin{tikzpicture}[scale=1]
        \foreach \y [count=\n] in {
            {4.65,2.94,3.45,3.19,4.73,3.80},
            {3.82,3.17,3.88,3.73,4.10,3.73},
            {4.71,3.51,4.21,4.64,4.37,4.29},
            {4.86,2.73,3.73,3.71,4.94,4.00},
            {4.61,2.88,3.94,4.08,4.88,4.09},
        } {
            \foreach \x [count=\m] in \y {
                \pgfmathsetmacro{\scaledx}{100 - (\x - 2.5) * 40}
                \pgfmathsetmacro{\xpos}{\m + (\m==6 ? 0.1 : 0)}
                \pgfmathsetmacro{\ypos}{-\n*0.5 - 0.2}
                \ifnum\n=6\else
                    \node[fill=LightGreen!\scaledx!DarkGreen, minimum height=5mm,minimum width=10mm, text=white] at (\xpos,\ypos) {\x};
                \fi
            }
        }
    
        % row labels
        \foreach \a [count=\i] in {airplane,car,smartphone,cell tower,pacemaker,\textbf{total}} {
            \pgfmathsetmacro{\xpos}{\i + (\i==6 ? 0.1 : 0)}
            \node[minimum height=5mm,minimum width=10mm,left,right,rotate=45] at (\xpos-0.1, -0.3) {\a};
        }
    
        % column labels
        \foreach \a [count=\i] in {accountability,ethical standards,cyber security,safety, trustworthiness} {
            \pgfmathsetmacro{\ypos}{-\i*0.5 - 0.2}
            \node[minimum height=5mm,minimum width=10mm,left] at (0.4,\ypos) {\a};
        }
    \end{tikzpicture}
    \caption{Participants' mean importance ratings of our desiderata in the context of the considered settings.}
    \label{xhw_study::fig::importance_desiderata}
    \Description{The figure shows a six-by-five heatmap comprising decimal numbers between 1 (lowest) and 5 (highest) for each entry in the six-by-five grid. On the x-axis, the settings "airplane", "car", "smartphone", "cell tower", and "pacemaker" are displayed together with an additional column labeled "total" that is separated from the heatmap by some whitespace. On the y-axis, the five desiderata "accountability", "ethical standards", "cyber security", "safety", and "trustworthiness" are given. The heat map entries are colored in different shades of blue depending on their value. The higher the value, the darker the blue color.}
\end{figure}

\subsubsection{Importance of Desiderata in Different Settings}
\label{xhw_study::subsubsec::desiderata_importance}
We asked participants about their perceived importance of five desiderata on a scale from \textit{1---not at all important} to \textit{5---extremely important}. 
The most important desiderata were cyber security ($M$$=$$4.29$, $SD$$=$$1.11$) and trustworthiness ($M$$=$$4.09$, $SD$$=$$1.19$). 
Safety comes in third ($M$$=$ $4.00$, $SD$$=$$1.31$) followed by accountability ($M$$=$$3.80$, $SD$$=$$1.35$) and ethical standards ($M$$=$$3.73$, $SD$$=$$1.28$).
\autoref{xhw_study::fig::importance_desiderata} reports more fine-grained mean values, connecting each desideratum to the different microchip application settings.

\subsubsection{Importance of Information Types}
We asked participants to rate the importance of different types of information in relation to the deployment setting and desideratum respectively, on a scale from \textit{1---not at all important} to \textit{5---extremely important}. 
Across all desiderata and settings, information on \textit{which functionality} a microchip provides was rated the most important ($M$$=$$3.60$, $SD$$=$$1.33$).
This is followed by information about how a microchip was \textit{approved for use} ($M$$=$$3.56$, $SD$$=$$1.36$) and how the microchip \textit{interacts with the surrounding system} ($M$$=$$3.51$, $SD$$=$$1.38$).
Participants placed less importance on the manufacturing aspects, namely information on \textit{who manufactured} the microchip ($M$$=$$3.32$, $SD$$=$$1.38$) 
and information on \textit{how a microchip was manufactured} ($M$$=$$3.25$, $SD$$=$$1.35$).

\pgfplotstableread{
Label                t1 t2 t3 t4 t5
who~manufactured     26 29 52 55 88
how~interacts        23 22 44 59 102
how~approved         19 22 38 59 112
which~functionality  20 29 39 61 101
how~manufactured     23 37 57 52 81
}\infsetairplane

\pgfplotstableread{
Label                t1 t2 t3 t4 t5
who~manufactured     49 55 64 43 39
how~interacts        48 42 56 54 50
how~approved         50 48 51 48 53
which~functionality  36 39 69 48 58
how~manufactured     53 58 57 47 35
}\infsetcar

\pgfplotstableread{
Label                t1 t2 t3 t4 t5
who~manufactured     38 35 61 60 56
how~interacts        36 32 56 59 67
how~approved         33 32 58 62 65
which~functionality  32 29 56 56 77
how~manufactured     45 39 69 50 47
}\infsetsmartphone

\pgfplotstableread{
Label                t1 t2 t3 t4 t5
who~manufactured     44 42 60 50 54
how~interacts        38 37 55 57 63
how~approved         31 36 57 64 62
which~functionality  34 29 58 70 59
how~manufactured     35 54 73 43 45
}\infsetcelltower

\pgfplotstableread{
Label                t1 t2 t3 t4 t5
who~manufactured     20 25 38 73 94
how~interacts        15 12 37 66 120
how~approved         12 12 30 66 130
which~functionality  9 12 34 64 131
how~manufactured     14 25 45 71 95
}\infsetpacemaker

\begin{figure*}[htbp]
    \centering
    \begin{tikzpicture}
        \begin{groupplot}[group style={group size=2 by 3, horizontal sep=2.4cm, vertical sep=0.9cm},
        /pgfplots/legend style={at={(0.5,4.6)}, anchor=north, legend columns=5}]
            \pgfplotsset{my bar/.style={width=6cm,
                xbar stacked, ymin=0,  
                xmin=0,
                y=10pt,
                bar width=5pt,
                ytick=data, 
                yticklabels from table={\infsetairplane}{Label},
                enlarge y limits={abs=5pt},
                axis y line*=left,
                axis x line*=bottom,
                xtick align=outside,
                ytick align=inside,
                font=\footnotesize,
                enlarge x limits={abs=5pt},
                title style={yshift=-0.25cm}
                }}
            \nextgroupplot[my bar, title=\textbf{steering of airplane}]
                \addplot [fill=c1] table [x=t1, meta=Label,y expr=\coordindex] {\infsetairplane};
                \addplot [fill=c2] table [x=t2, meta=Label,y expr=\coordindex] {\infsetairplane};
                \addplot [fill=c3] table [x=t3, meta=Label,y expr=\coordindex] {\infsetairplane};
                \addplot [fill=c4] table [x=t4, meta=Label,y expr=\coordindex] {\infsetairplane};
                \addplot [fill=c5] table [x=t5, meta=Label,y expr=\coordindex] {\infsetairplane};
            \nextgroupplot[my bar, title=\textbf{entertainment system in car}]
                \addplot [fill=c1] table [x=t1, meta=Label,y expr=\coordindex] {\infsetcar};
                \addplot [fill=c2] table [x=t2, meta=Label,y expr=\coordindex] {\infsetcar};
                \addplot [fill=c3] table [x=t3, meta=Label,y expr=\coordindex] {\infsetcar};
                \addplot [fill=c4] table [x=t4, meta=Label,y expr=\coordindex] {\infsetcar};
                \addplot [fill=c5] table [x=t5, meta=Label,y expr=\coordindex] {\infsetcar};
            \nextgroupplot[my bar, title=\textbf{fingerprint unlocking in smartphone}]
                \addplot [fill=c1] table [x=t1, meta=Label,y expr=\coordindex] {\infsetsmartphone};
                \addplot [fill=c2] table [x=t2, meta=Label,y expr=\coordindex] {\infsetsmartphone};
                \addplot [fill=c3] table [x=t3, meta=Label,y expr=\coordindex] {\infsetsmartphone};
                \addplot [fill=c4] table [x=t4, meta=Label,y expr=\coordindex] {\infsetsmartphone};
                \addplot [fill=c5] table [x=t5, meta=Label,y expr=\coordindex] {\infsetsmartphone};
            \nextgroupplot[my bar, title=\textbf{wireless communication of cell tower}]
                \addplot [fill=c1] table [x=t1, meta=Label,y expr=\coordindex] {\infsetcelltower};
                \addplot [fill=c2] table [x=t2, meta=Label,y expr=\coordindex] {\infsetcelltower};
                \addplot [fill=c3] table [x=t3, meta=Label,y expr=\coordindex] {\infsetcelltower};
                \addplot [fill=c4] table [x=t4, meta=Label,y expr=\coordindex] {\infsetcelltower};
                \addplot [fill=c5] table [x=t5, meta=Label,y expr=\coordindex] {\infsetcelltower};
            \nextgroupplot[my bar, title=\textbf{heartbeat in pacemaker}, at = {($($(group c1r2.south west) + (0,-1.8cm)$ )!0.5!(group c2r2.south east)$) }]
                \addplot [fill=c1] table [x=t1, meta=Label,y expr=\coordindex] {\infsetpacemaker};
                \addplot [fill=c2] table [x=t2, meta=Label,y expr=\coordindex] {\infsetpacemaker};
                \addplot [fill=c3] table [x=t3, meta=Label,y expr=\coordindex] {\infsetpacemaker};
                \addplot [fill=c4] table [x=t4, meta=Label,y expr=\coordindex] {\infsetpacemaker};
                \addplot [fill=c5] table [x=t5, meta=Label,y expr=\coordindex] {\infsetpacemaker};
        \legend{1---not at all important,2---slightly important,3---moderately important,4---very important,5---extremely important}
        \end{groupplot}
     \end{tikzpicture}
    \caption{Importance of different types of information depending on the setting in which microchips are employed. The results are aggregated across all desiderata.}
    \label{xhw_study::fig::importance_information_settings}
    \Description{The figure shows five horizontal bar charts, every one of them comprising a title and five bars of five sections each. The bar charts are arranged in a V-shape: two on top, two in the middle, and a last one centered below the middle two. All bar charts share a common legend among them, which is placed on top of them all. The labels of each bar chart are the same apart from their title. From left to right, top to bottom, the bar chart titles read "steering of airplane", "entertainment system in car", "fingerprint unlocking in smartphone", "wireless communication of cell tower", and "heartbeat in pacemaker". On the x-axis of each chart, all bars span from 0 to 250. On the y-axis of each chart, the first bar (the top one) is labeled "how manufactured", the second bar is labeled "which functionality", the third bar is labeled "how approved", the fourth bar is labeled "how interacts", the fifth (and last) bar is labeled "who manufactured". The legend on at the top then assigns meanings to the segments of each bar. It comprises five labels and is organized in a single row. The first label says "1-not at all important", the second "2-slightly important", the third "3-moderately important", the fourth "4-very important", and the fifth "5-extremely important".}
\end{figure*}
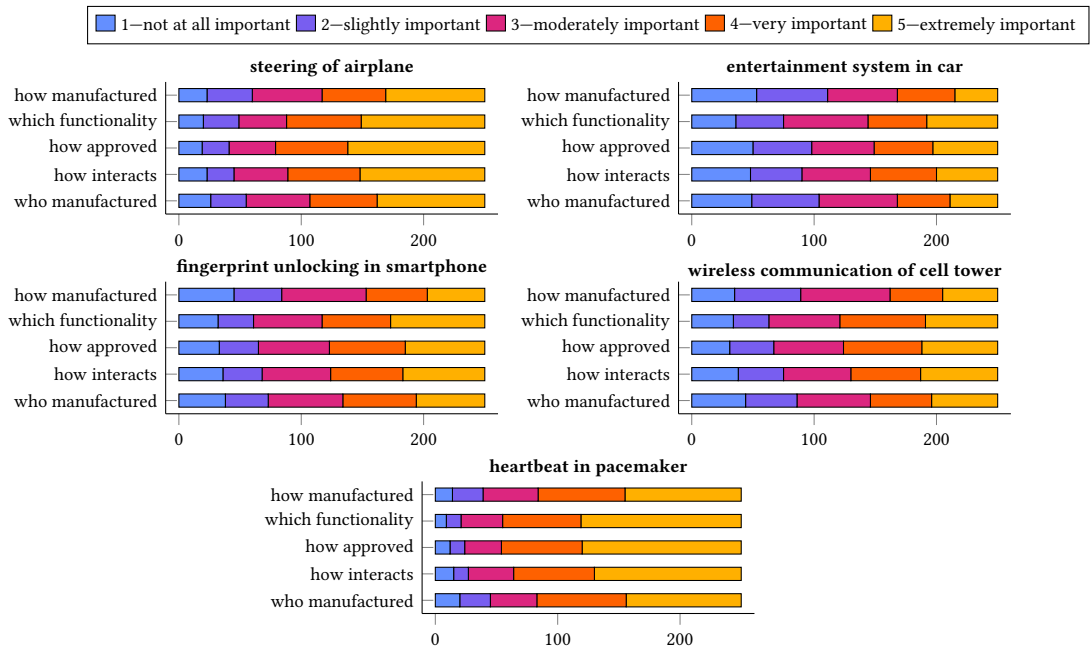

\pgfplotstableread{
Label                t1 t2 t3 t4 t5
who~manufactured     43 36 69 49 53
how~interacts        32 30 53 63 72
how~approved         29 26 45 62 88
which~functionality  24 36 51 58 81
how~manufactured     38 44 71 41 56
}\infdessafety

\pgfplotstableread{
Label                t1 t2 t3 t4 t5
who~manufactured     31 43 43 51 82
how~interacts        32 33 46 56 83
how~approved         28 36 52 60 74
which~functionality  27 25 65 53 80
how~manufactured     34 47 57 52 60
}\infdesaccountability

\pgfplotstableread{
Label                t1 t2 t3 t4 t5
who~manufactured     31 29 54 63 73
how~interacts        46 34 50 50 70
how~approved         34 31 47 59 79
which~functionality  40 29 48 59 74
how~manufactured     28 36 52 58 76
}\infdesethical

\pgfplotstableread{
Label                t1 t2 t3 t4 t5
who~manufactured     32 35 56 62 65
how~interacts        23 18 43 67 99
how~approved         22 26 42 62 98
which~functionality  18 27 39 67 99
how~manufactured     34 43 57 60 56
}\infdessecurity

\pgfplotstableread{
Label                t1 t2 t3 t4 t5
who~manufactured     40 43 53 56 58
how~interacts        27 30 56 59 78
how~approved         32 31 48 56 83
which~functionality  22 21 53 62 92
how~manufactured     36 43 64 52 55
}\infdestrustworthiness

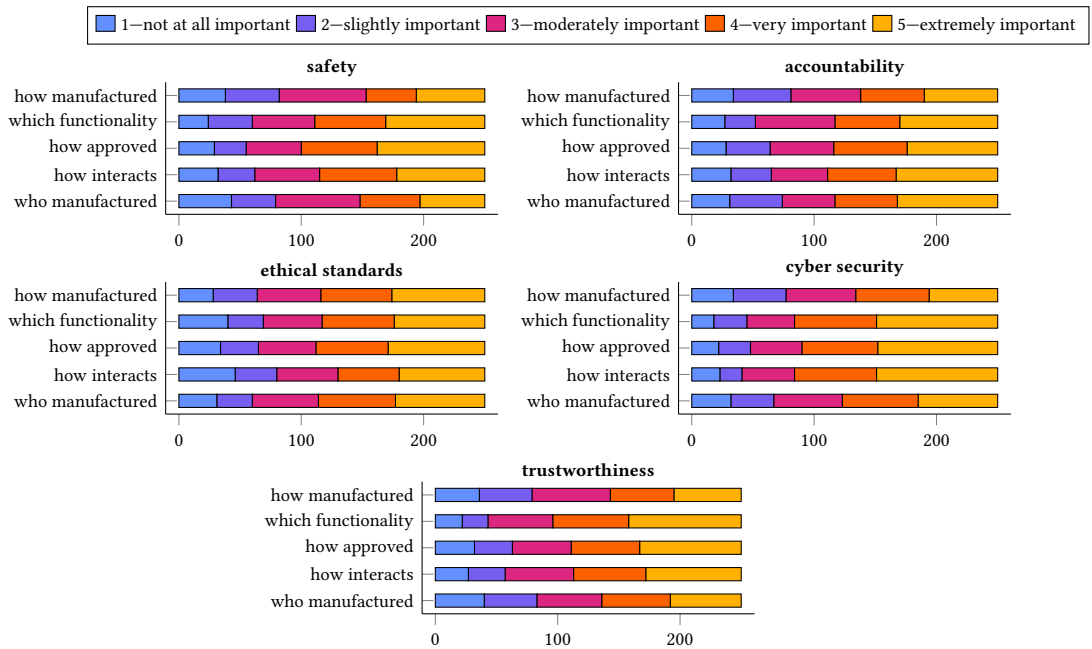
\begin{figure*}[htbp]
    \centering
    \begin{tikzpicture}
        \begin{groupplot}[group style={group size=2 by 3, horizontal sep=2.4cm, vertical sep=0.9cm},
        /pgfplots/legend style={at={(0.5,4.6)}, anchor=north, legend columns=5}]
            \pgfplotsset{my bar/.style={width=6cm,
                xbar stacked, ymin=0,  
                xmin=0,
                y=10pt,
                bar width=5pt,
                ytick=data, 
                yticklabels from table={\infdessafety}{Label},
                enlarge y limits={abs=5pt},
                axis y line*=left,
                axis x line*=bottom,
                xtick align=outside,
                ytick align=inside,
                font=\footnotesize,
                enlarge x limits={abs=5pt},
                title style={yshift=-0.25cm}
                }}
            \nextgroupplot[my bar, title=\textbf{safety}]
                \addplot [fill=c1] table [x=t1, meta=Label,y expr=\coordindex] {\infdessafety};
                \addplot [fill=c2] table [x=t2, meta=Label,y expr=\coordindex] {\infdessafety};
                \addplot [fill=c3] table [x=t3, meta=Label,y expr=\coordindex] {\infdessafety};
                \addplot [fill=c4] table [x=t4, meta=Label,y expr=\coordindex] {\infdessafety};
                \addplot [fill=c5] table [x=t5, meta=Label,y expr=\coordindex] {\infdessafety};
            \nextgroupplot[my bar, title=\textbf{accountability}]
                \addplot [fill=c1] table [x=t1, meta=Label,y expr=\coordindex] {\infdesaccountability};
                \addplot [fill=c2] table [x=t2, meta=Label,y expr=\coordindex] {\infdesaccountability};
                \addplot [fill=c3] table [x=t3, meta=Label,y expr=\coordindex] {\infdesaccountability};
                \addplot [fill=c4] table [x=t4, meta=Label,y expr=\coordindex] {\infdesaccountability};
                \addplot [fill=c5] table [x=t5, meta=Label,y expr=\coordindex] {\infdesaccountability};
            \nextgroupplot[my bar, title=\textbf{ethical standards}]
                \addplot [fill=c1] table [x=t1, meta=Label,y expr=\coordindex] {\infdesethical};
                \addplot [fill=c2] table [x=t2, meta=Label,y expr=\coordindex] {\infdesethical};
                \addplot [fill=c3] table [x=t3, meta=Label,y expr=\coordindex] {\infdesethical};
                \addplot [fill=c4] table [x=t4, meta=Label,y expr=\coordindex] {\infdesethical};
                \addplot [fill=c5] table [x=t5, meta=Label,y expr=\coordindex] {\infdesethical};
            \nextgroupplot[my bar, title=\textbf{cyber security}]
                \addplot [fill=c1] table [x=t1, meta=Label,y expr=\coordindex] {\infdessecurity};
                \addplot [fill=c2] table [x=t2, meta=Label,y expr=\coordindex] {\infdessecurity};
                \addplot [fill=c3] table [x=t3, meta=Label,y expr=\coordindex] {\infdessecurity};
                \addplot [fill=c4] table [x=t4, meta=Label,y expr=\coordindex] {\infdessecurity};
                \addplot [fill=c5] table [x=t5, meta=Label,y expr=\coordindex] {\infdessecurity};
            \nextgroupplot[my bar, title=\textbf{trustworthiness}, at = {($($(group c1r2.south west) + (0,-1.8cm)$ )!0.5!(group c2r2.south east)$) }]
                \addplot [fill=c1] table [x=t1, meta=Label,y expr=\coordindex] {\infdestrustworthiness};
                \addplot [fill=c2] table [x=t2, meta=Label,y expr=\coordindex] {\infdestrustworthiness};
                \addplot [fill=c3] table [x=t3, meta=Label,y expr=\coordindex] {\infdestrustworthiness};
                \addplot [fill=c4] table [x=t4, meta=Label,y expr=\coordindex] {\infdestrustworthiness};
                \addplot [fill=c5] table [x=t5, meta=Label,y expr=\coordindex] {\infdestrustworthiness};
        \legend{1---not at all important,2---slightly important,3---moderately important,4---very important,5---extremely important}
        \end{groupplot}
     \end{tikzpicture}
    \caption{Importance of different kinds of information depending on the desiderata to be evaluated by the end user. The results are aggregated across all settings.}
    \label{xhw_study::fig::importance_information_desiderata}
    \Description{The figure shows five horizontal bar charts, every one of them comprising a title and five bars of five sections each. The bar charts are arranged in a V-shape: two on top, two in the middle, and a last one centered below the middle two. All bar charts share a common legend among them, which is placed on top of them all. The labels of each bar chart are the same apart from their title. From left to right, top to bottom, the bar chart titles read "safety", "accountability", "ethical standards", "cyber security", and "trustworthiness". On the x-axis of each chart, all bars span from 0 to 250. On the y-axis of each chart, the first bar (the top one) is labeled "how manufactured", the second bar is labeled "which functionality", the third bar is labeled "how approved", the fourth bar is labeled "how interacts", the fifth (and last) bar is labeled "who manufactured". The legend on at the top then assigns meanings to the segments of each bar. It comprises five labels and is organized in a single row. The first label says "1-not at all important", the second "2-slightly important", the third "3-moderately important", the fourth "4-very important", and the fifth "5-extremely important".}
\end{figure*}

\autoref{xhw_study::fig::importance_information_settings} shows the trend that the type of information desired by end users depends on the application setting in which they are used at least to some extent. Information on the functionality of a microchip, how it has been approved for use, and how it interacts with the system were perceived to be more important than the other types of information, particularly in airplane and pacemaker settings.
For other settings, such as the car and the smartphone, these differences still exist but are not as pronounced. For instance, information on a microchip's functionality was rated the most important for the smartphone setting.

\autoref{xhw_study::fig::importance_information_desiderata} shows that the desired information types may also depend on the target desideratum.
For example, to evaluate cyber security and trustworthiness, information on the microchip's functionality, how it has been approved for use, and how it interacts with the system were rated more important than the manufacturing-related information.
A similar trend was observed for safety, although here, information on how the microchips have been approved has a small edge over the two others.
When end users want to evaluate accountability or ethical standards, the ratings were similar and no particular types of information stood out.

\begin{tcolorbox}
\subsubsection*{Summary in Light of RQ2} 
Participants had diverse perceptions regarding the criticality of different application settings for microchips. 
While participants considered all five desiderata very important ($M$$>$$3.5$ for all), cyber security and trustworthiness emerged to be the more important ones.
For information types, participants desired to know more about the microchip's functionality, how it is approved for use, and how it interacts with the underlying system than about the manufacturing processes. 
We also observe the trend that the desired information types depend on the application setting and the target desideratum, and we quantitatively test the correlations in~\autoref{xhw_study::subsec::regression_information}.
\end{tcolorbox}

\subsection{RQ3: Factors Shaping End Users' Information Needs}
\label{xhw_study::subsec::regression_information}

To gain more granular insight into the factors that shape end users' information needs, we applied multilevel regression modeling to each of the five information types.
For settings, the \textit{microchips controlling the entertainment system in a car} was used as a baseline because of its lowest criticality rating (see~\autoref{xhw_study::subsubsec::setting_criticality}).
For desiderata, \textit{ethical standards} was chosen as the baseline as participants rated it as least important (see \autoref{xhw_study::subsubsec::desiderata_importance}).
Compared to the intercept-only models without any predictors, the random intercept models containing the settings and desiderata had significantly lower deviances (\eg, $\chi^2(8)$$=$$59.95$, $p$$<$$.001$ for the \textit{which functionality} model), indicating a better fit to our data.

\begin{table*}[htbp]
    \centering
    \footnotesize
    \caption{Multilevel regression analysis based on participants' ratings of the importance of receiving different types of information to evaluate a desideratum in a given setting, on a scale from \textit{1---not at all important} to \textit{5---extremely important}.}% (see \autoref{xhw_study::question::v0_information} for an example question presented to participants).}
    \label{xhw_study::tab::regression_information}
    \begin{tabularx}{\textwidth}{XL{1.5cm}L{1.5cm}L{1.5cm}L{1.5cm}L{1.5cm}}
        \toprule
         & \textbf{which func-} & \textbf{how} & \textbf{how} & \textbf{who manu-} &  \textbf{how manu-} \\
         & \textbf{tionality} &  \textbf{interacts} & \textbf{approved}  & \textbf{factured}  & \textbf{factured}\\
         \textit{Predictors} & \textit{Est.} & \textit{Est.} & \textit{Est.} &  \textit{Est.} & \textit{Est.} \\
         \midrule
        intercept: ethical standards (desideratum) $\times$ car (setting) & \textbf{2.61***} & \textbf{2.35***} & \textbf{2.63***} & \textbf{2.66***}  & \textbf{2.71***} \\
        \midrule
         \textit{setting (baseline=car)} & & & & & \\
        smartphone & 0.25 %**
        & 0.29
        %**
        & \textbf{0.35***} &  0.25 %**
        & \textbf{0.37***} \\
        cell tower & 0.15 & 0.21 %*
        & \textbf{0.34***} & 0.22 %**
        & 0.24 
        %**
        \\
        pacemaker & \textbf{0.97***} & \textbf{0.99***} &  \textbf{1.13***} & \textbf{1.02***} & \textbf{0.92***} \\
        airplane & \textbf{0.56***} & \textbf{0.71***} &  \textbf{0.87***} & \textbf{0.72***} & \textbf{0.73***} \\
        \midrule
        \textit{desideratum (baseline=ethical standards)} & & & & & \\
        accountability & 0.13 & 0.23 %*
        & -0.02 & -0.25 %**  
        & -0.04 \\
        safety & 0.14 & 0.18 %*
        & 0.13 &  \textbf{-0.36***} & \textbf{-0.35***} \\
        trustworthiness & \textbf{0.32***} & 0.25 %**
        & 0.02 & \textbf{-0.30***} & -0.29 %**
        \\
        cyber security & \textbf{0.41***} & \textbf{0.54***} &  0.27 %** 
        & -0.24 %** 
        & -0.11 \\
        \midrule
        desire to understand more about microchips & \textbf{0.53***} & \textbf{0.63***} & 0.41 %**
        & 0.50 %**
        & 0.43 %* 
        \\
        ATI score & \textbf{0.26***} & \textbf{0.27***} & 0.23 %**
        & 0.26 %**
        & 0.24 %**
        \\
        % \midrule
        \midrule
        marginal $R^2$    & 0.171 & 0.190 & 0.155 & 0.165 & 0.129\\
        conditional $R^2$ & 0.461 & 0.469 & 0.507 & 0.554 & 0.521\\    
        \bottomrule
     \end{tabularx}
     \textit{
     % * p < 0.05; ** p < 0.01, 
     *** $p$$<$$.001$; we only highlighted coefficients with $p$$<$$.001$ due to the Bonferroni correction}
     \label{xhw_study::subsec::regression_results}
     \Description{The table is organized in six columns. In the first column on the left, the predictors for multilevel regression analysis are given. Each of the other five major columns provides details on a type of information about microchips, starting with "which functionality" on the left and ending with "how manufactured" on the right. Each row of the five major columns gives the Estimate ("Est.") for one predictor. Here, "***" following an estimate means p<0.001, and respective entries are highlighted in boldface. In the first line below the table headers, the intercept of regression analysis is given for each kind of information. Next, a sub-header spanning the entire table titled "settings (baseline=car)" opens the section with the analysis results for the four settings "smartphone", "cell tower", "pacemaker", and "airplane". The estimates for these settings are given in the next 4 rows. Next, a sub-header spanning the entire table titled "Desiderata (baseline=ethical standards)" opens the section with the analysis results for the four desiderata "accountability", "safety", "trustworthiness", and "cyber security". The estimates for these desiderata are given in the next 4 rows. In the next section, the first row is titled "desire to understand more about microchips" and lists respective estimates. The second and final row of this section is titled "ATI score" and lists respective estimates. Finally, in the last section of the table, the first row lists results for "marginal R^2" and the second row "conditional R^2".}
\end{table*}

We then tested the specific predictors for the significant explanatory power expressed by marginal $R^2$.
In addition to the application setting's perceived criticality and the desideratum's perceived importance as the main effects, we included participants' general desire to understand more about microchips as a binary predictor (\textit{no} as the baseline) and their \ac{ATI} score~\cite{franke2019personal} (applying grand-mean centering, utilizing the \ac{ATI} mean value as the baseline).
We further tried including participants' demographics (\ie, gender, age, and educational background) in our models, but they did not add significantly more explanatory power.
We thus omit these variables from the analysis.
Our final models with the random intercept reached a better fit ($\ac{AIC}$$=$$[3725.2 - 3913.7]$;  $\ac{BIC}$$=$$[3806.9 - 4062.5]$) compared to our base models ($\ac{AIC}$$=$$[3969.6 - 4123.4]$;  $\ac{BIC}$$=$$[3985.0 - 4138.8]$).

\autoref{xhw_study::subsec::regression_results} shows the regression outputs. 
Below we unpack a few key findings. 
We report separate regression models that look into the interaction effects between settings and desiderata in \autoref{xhw_study::app::regression_results}, which show similar patterns as the findings reported here.\footnote{While the models with interaction effects enable a more nuanced examination of information needs in response to individual vignettes, the number of participants per vignette was limited to 25, negatively affecting the statistical power and increasing the alpha error accumulation. 
We thus opted to report the interaction effects in \autoref{xhw_study::app::regression_results}.}

\paragraph{Higher Information Needs for Critical Settings} 
Looking at the main effect of application settings, we observe that settings with higher criticality ratings were significantly correlated with higher information needs. 
Compared to \textit{car} as the baseline, participants gave significantly higher ratings across the five information types for vignettes featuring \textit{airplane} and \textit{pacemaker}. 
\textit{Smartphone} and \textit{cell tower} as setting also drove up the information needs to some degree. 
However, this pattern only applies to certain types of information, namely \textit{how the microchip is approved for use} (for both settings) and \textit{how the microchip is manufactured} (only for \textit{smartphone}).

\paragraph{Nuanced Influences from Desiderata}
In contrast to findings on application settings, where there was a clear association between high perceived criticality and high information needs, the effect of desiderata on information needs is more nuanced.
Compared to ethical standards, participants had a stronger desire for information on the microchip's functionality and how it interacts with the system when \textit{cyber security} was the target desideratum. 
Participants also valued information on the microchip's functionality for \textit{trustworthiness}. 
Conversely, participants attached less importance to information about the microchip's manufacturing process, but only when the target desiderata were \textit{safety} (for both information types) and \textit{trustworthiness} (for \textit{who manufactured} only).
Between ethical standards and \textit{accountability}, participants' information needs were similar with no statistically significant differences.

\paragraph{Information Needs Shaped by the Desire to Understand and \acs{ATI}}
A general desire of participants to know more about microchips also shapes participants' information needs. 
The coefficients are positive across the five information types, and the influences were particularly pronounced for information on the microchip's functionality and how the microchip interacts with the system.
These observations are in line with our findings regarding RQ1, where participants shared their willingness to learn more about microchips open-endedly, and their existing understandings revolve around functionalities and application settings.
Similarly, a higher \ac{ATI} score contributes to more desire for information, particularly for functionality and interactions with the system. 

\begin{tcolorbox}
\subsubsection*{Summary in Light of RQ3.}
The factors driving end users' information needs are multifaceted. 
Higher information needs generally occurred when participants perceived the setting in which the microchip was deployed to be highly critical, whereas desiderata do not consistently predict information needs. 
Participants' general desire to understand microchips and \ac{ATI} also played a significant role in shaping information needs, particularly for the microchip's functionality and how it interacts with the system.
\end{tcolorbox}
\section{Discussion}
\label{xhw_study::sec::discussion}

Below, we reflect on the fundamental questions of why end users need to understand more about microchips and the role of end users in the microchip ecosystem (\autoref{sec:why:understanding}). 
We then discuss our findings' implications for future research that promotes user understanding of microchips and microchip transparency (\autoref{sec:hci:implications}). 
Finally, we reflect on our work's policy implications considering regulatory efforts around microchips (\autoref{sec:policy:implications}).

\subsection{Do End Users Need to Understand More About Microchips?}
\label{sec:why:understanding}
Our study is motivated by the fact that microchips run the electronics of the world and are featured prominently in regulatory efforts, yet microchips remain largely opaque from the general public view.
Nonetheless, they play an increasingly vital role in security as they often form the root of trust in a system, \eg, as a cryptographic accelerator, \ac{HSM}, or \ac{TEE}.
In other domains and application areas, such as \acs{AI} and \ac{IoT} devices, we have seen concrete evidence that a lack of transparency causes security and trust issues~\cite{von2021transparency}. 
In contrast, end users are empowered to make more informed decisions with a better understanding of the system's inner structure and potential risks~\cite{emami2020ask,johnson2020impact}. 
Thus, we see the value of at least envisioning the integration of end users into the hardware ecosystem since their role is largely overlooked at the moment.
Our findings further underscore the necessity of helping end users understand more about microchips in a few ways. 

First, the need is supported by our participants' own preferences---76\% of participants indicated they would personally like to understand more about microchips, recognizing microchips' omnipresence in their daily lives and expressing particular interest in knowing more about microchip's functionality and interactions with the underlying system. 

Second, while our participants exhibited a basic understanding of what microchips are and where they are used, we found a lack of awareness regarding potential security and privacy concerns, the critical societal aspects of microchips, and misconceptions such as linking microchips to animal implants and conspiracy theories. 
Beliefs in such conspiracies can lead to hesitations in adopting new technologies and mistrust in government bodies.
In fact, during the COVID-19 pandemic, conspiracy theorists falsely claimed vaccines were used to implant microchips into people, which led to lower vaccination rates~\cite{ROMER2020113356,ULLAH202193}.

Finally, helping end users better understand microchips has numerous practical impacts. 
For instance, with a better understanding, end users would be more equipped to participate in discussions around legislative efforts such as the US CHIPS and Science Act~\cite{uschips2022} and the European Chips Act~\cite{euchips2022}, and---as citizens in democratic societies---to inform and hold their governments accountable for the significant investment through such programs.

%Another related avenue is the \enquote{right to repair} directive recently adopted by the European Parliament to ensure that manufacturers provide timely and cost-effective repair services and inform consumers about their rights to repair~\cite{EuroParl24}. 
%With more knowledge of how microchips work, end users can more meaningfully take advantage of the right to repair, such as by identifying suitable replacement parts and finding the right documentation, when devices and infrastructure involving microchips break down~\cite{svensson2018emerging,DBLP:conf/cscw/RosnerA14}. 

With all the reasons summarized, we believe that the remaining question is not \textit{whether} we need to help end users understand more about microchips, but rather \textit{when} and \textit{how} to achieve this objective. 
Regarding the \enquote{when} aspect, it is important to acknowledge that end users have varying degrees of decision-making across the different scenarios in which microchips are deployed. 
For instance, when the device in question is a computer or tablet provided with dozens of microchips, we can reasonably expect that end users may adjust their level of trust in the device to purchase based on information about the microchip's performance (\eg., about its functionality), security (\eg, based on certifications), and ethical considerations (such as fair wages and working conditions) for workers involved in the manufacturing processes. 

However, this is less likely the case when deciding which airplane to take, as microchips are deployed en masse in planes and are generally inaccessible to end users.
Here, other factors such as the ticket's price and availability come as priorities~\cite{anwar2021effect}, and end users can only rarely choose the airplane type. 
Interestingly, this stands in contrast to our finding that our participants rated the airplane and pacemaker scenarios as the most critical (and more critical scenarios drive higher information needs).
In the pacemaker scenario, the deployment of microchips is less complicated.
Beyond medical reasons~\cite{doi:10.1161/01.CIR.91.4.1063}, end users have a fair degree of decision-making agency between individual devices and vendors.

The key to finding the right \enquote{when} moment is to identify other application settings that are not only important and relevant to end users, but also offer space for end users to make meaningful and informed decisions.
Going beyond the scenarios presented in our survey, we could imagine smartwatches and smartglasses as well as \acs{IoT} and smart home devices to fall into this category.
However, other complex applications, such as industrial machines and (digital) infrastructure components, are likely out of scope.

\subsection{Towards Microchip Transparency for End Users}
\label{sec:hci:implications}
We believe that future interdisciplinary research is required and the usable security community is uniquely positioned to tackle the \enquote{how} aspect of helping end users better understand microchips.
Below, we outline a few possible directions informed by our findings and speculate potential ideas to explore based on our own knowledge. 

\subsubsection{Building Mental Models of Microchips}
As our study is first-of-its-kind for the topic and exploratory in nature, we gauged participants' understanding of microchips in a simple open-ended question.
Our initial results pave the way for more thorough analyses of end users' mental models of microchips, which serve as foundational knowledge for any tools, resources, and educational interventions that seek to teach users about microchips.
For instance, future work can elicit end users' mental models in qualitative methods such as interviews, focus groups, co-design sessions, and drawing activities that enable deeper insights into users' reasoning processes and why misconceptions occur~\cite{jones2011mental}. 

Future work can also replicate prior studies on the mental models of computer security~\cite{wash2010folk,camp2009mental} and privacy~\cite{DBLP:journals/popets/OatesAMSZBC18} in the microchip setting to see to what extent users' existing models and metaphors still apply. 
Moreover, as prior work has consistently demonstrated the gaps between experts and laypeople regarding mental models~\cite{DBLP:journals/popets/OatesAMSZBC18,camp2007experimental,DBLP:conf/uss/BinkhorstFKPL22}, and microchips remain opaque even to experts~\cite{DBLP:conf/re/SpeithSBZBP24}, it is crucial to compare the mental models held by non-expert end users with those held by other stakeholders in the hardware ecosystem (such as designers, manufacturers, system integrators, and policymakers)~\cite{DBLP:conf/re/SpeithSBZBP24} in order to identify and close the gaps.

\subsubsection{Deciding Specific Information to Provide to End Users}
\label{subsubsec:info-to-provide}
Our study hints at the types of information that end users prioritize for understanding more about microchips.
However, the categories we presented in our study were quite broad. 
Future work is needed to empirically compare the effectiveness and downstream impacts on users (\eg, in terms of comprehension, trust in the system, and purchase behaviors) across the different information types, ideally with vignettes that feature the specific information adapted for the application setting.
Inspirations can also be drawn from the nudging literature for the framing of the presented information~\cite{acquisti2017nudges}. 

For instance, since our findings demonstrate that end users may lack awareness of the broader societal, economic, and security implications of microchips regarding risks and harms, future work can explore the effectiveness of presenting information that saliently features concrete harms.
Examples of harm can include hardware security issues, critical malfunctions in pacemakers, and environmental harms in communities involved in the mining of resources required for microchip manufacturing.
By making more informed purchase decisions, collective actions from end users could help improve working conditions and reduce environmental impact.

\subsubsection{Designing and Evaluating Transparency Mechanisms for Microchips}
Once the specific information to be provided has been determined, the follow-up question is how to effectively convey the information to laypeople through transparency mechanisms specifically applicable to microchips.
For instance, hardware datasheets have existed for a while. 
They contain information on the functionality and connectivity of a microchip as well as on its ideal operating conditions. 
However, they often contain technical jargon that makes them inaccessible to end users.
Drawing from standardized labels for \acs{IoT} devices~\cite{emami2020ask} and mobile apps~\cite{Cranor2022Mobile}, model cards for ML models~\cite{DBLP:conf/fat/MitchellWZBVHSR19}, and datasheets for datasets~\cite{DBLP:journals/cacm/GebruMVVWDC21}, we see the promise of creating \enquote{microchip labels} that enhance existing hardware datasheets beyond providing the typical technical documentations to make them more accessible and useful to end users. 
Taking our findings into account, the label can cover the microchip's functionality, interaction with the system, supply chain actors, involved certification bodies, and more. 
Such a label for a tablet computer could, for example, provide a score related to all microchips in the device based on manufacturing location and conditions, sustainability, and security. 
A QR code as part of this label could then lead to a list of all contained microchips as well as details on properties such as their functionality, manufacturer, and interoperability.
Similar to the IoT label development pipeline~\cite{DBLP:journals/cacm/CranorAN24}, much more work is needed after the initial proposal to reach a consensus on details surrounding the label (\eg, having minimal vs. more complicated labels, the presence of a QR code, the label's size, and how the label is encouraged or mandated in regulations).

\subsection{Involving End Users in Regulatory Initiatives Around Microchips}
\label{sec:policy:implications}
Microchips represent a subject with natural policy implications. 
Against the background of public discourse about the use of Huawei equipment in network infrastructure~\cite{Webster2019} and the political efforts to promote domestic chip production in the United States and the European Union~\cite{euchips2022,uschips2022}, one of our key findings stands out---our participants were less interested in information about how and by whom a microchip was manufactured compared to the other types of information, whereas this aspect has been featured front and center in these regulatory initiatives.

Our study suggests that there is a potential gap between what legislators prioritize to address versus what end users desire to know. 
This may be due to the fact that microchip manufacturing is an intricate process that end users are mostly unaware of.
Given the level of knowledge required to comprehend microchip manufacturing, we argue that it would be best to leave technical manufacturing details to the regulators and instead focus on ethical aspects of manufacturing as well as microchip functionality and interaction within a system when designing explanations for end users.

One thing is known for sure: we cannot assume that the current multi-billion dollar investments from regulators will guarantee end-user trust in microchips. 
Therefore, similar to existing research on user perceptions of rights prescribed in the \ac{GDPR}~\cite{DBLP:conf/chi/KyiMSRB24,DBLP:conf/IEEEares/ManginiTM20,strycharz2020data,DBLP:conf/soups/KaushikYD021}, more work is needed to understand end users' perceptions of ongoing regulatory initiatives around microchips in order to capture and embed laypeople's opinions about microchips into policymaking. 

\section{Conclusion}
\label{xhw_study::sec::conclusion}
Microchips have become ubiquitous in people's daily lives, whether in the cars we drive, the phones we use, or even in our household appliances.
This observation highlights their indispensable role within socio-technical systems.
To better understand end-user perceptions of microchips, we conducted a survey with 250 participants.

While our participants appear to have a fundamental understanding of what microchips are and what they are used for, their knowledge of the consequences of microchip malfunction and their impact on society, in general, seems limited.
In particular, few participants had issues like cyber security, trustworthiness, or safety in mind, yet they considered them very important when explicitly asked about them.
Furthermore, our participants' information needs depend on their general affinity for technology, their willingness to understand more about microchips, and the considered desideratum and use case.
Based on our findings, future work could further explore end users' mental models of microchips and how to determine and convey information about them, so that end users can make more informed decisions about the purchase and use of electronic devices in the future.

\section*{Acknowledgments}
We thank an anonymous Prolific user whose answer to \autoref{xhw_study::question::perception} inspired our paper's title.

Work on this paper was funded by the Deutsche Forschungsgemeinschaft (DFG, German Research Foundation) under Germany's Excellence Strategy---\href{https://casa.rub.de}{EXC 2092 CASA}---390781972, through the DFG grant 389792660 as part of \href{https://perspicuous-computing.science}{TRR~248}, and by the Volkswagen Foundation grants AZ~9B830, AZ~98509, and AZ~98514 \href{https://explainable-intelligent.systems}{\enquote{Explainable Intelligent Systems}} (EIS). 

The Volkswagen Foundation and the DFG had no role in preparation, review, or approval of the manuscript; or the decision to submit the manuscript for publication. 
The authors declare no other financial interests.

\bibliographystyle{ACM-Reference-Format}
\bibliography{main}

%TC:ignore
\appendix
\section{Survey Material}
\label{xhw_study::app::survey}

\subsection{Your Perspective on Microchips}
\label{xhw_study::subapp::informed}
\begin{enumerate}
    \item \label{xhw_study::question::consent}
        Thank you for your interest in our study!
          
        \textbf{Purpose:} Increasing digitalization in all areas of life is being driven by the constantly rising performance and efficiency of microchips. With this survey, we would like to learn more about the desired goals of end users regarding their understanding of microchips, and how these goals may be achieved. By faithfully completing this survey, you can help make microchips more understandable to end users in the future.
        
        \textbf{Duration:} Participation in the study is expected to take a maximum of 25 minutes. You are not subject to any anticipated risks by participating. Please answer the survey as honestly as possible. You may stop at any time if you no longer wish to participate in the study. In case you drop out of the study, all responses recorded so far will be discarded.
          
        \textbf{Data Privacy Statement \& Informed Consent:} 
        Your responses to this study are stored in anonymized form in a way which will not reveal your identity. 
        No data will be passed on to third parties. 
        By starting this questionnaire you consent to data collection for the purposes of conducting this study. 
        Your personal data is processed based on Article 6 (1) a GDPR and \textit{[redacted for review]}. %§ 17 DSG NRW (Datenschutzgesetz Nordrhein-Westfalen (Data Protection Regulation North Rhine-Westphalia)). 
        You have the right to revoke your consent to the data processing at any time as well as to request information, correction, processing restrictions and deletion of the data stored about you. 
        To exercise these rights, please contact the email address listed below. 
        The responsible supervisory authority is the \textit{[redacted]}. %State Data Protection Officer of the State of North Rhine-Westphalia. 
        If you have additional questions about data protection, please contact \textit{[redacted]}. %dsk@mpi-sp.org.
        
        To participate, you must be 18 years of age or older and a resident of the United States.
    \begin{enumerate}
        \item I am 18 years of age or older. \mychoice{Yes} \mychoice{No}
        \item I am a resident of the United States. \mychoice{Yes} \mychoice{No}
        \item I confirm that I accept the participation conditions for this study. \mychoice{Yes} \mychoice{No}
        \item I do not agree and don't want to participate. \mychoice{Yes} \mychoice{No}
    \end{enumerate}
    \setcounter{qcounter}{\value{enumi}}
\end{enumerate}

\subsection{Your Interaction with Technical Systems}
\label{xhw_study::subapp::ati}
\begin{enumerate}
    \setcounter{enumi}{\value{qcounter}}
    \item \label{xhw_study::question::ati} 
        In the following questionnaire, we will ask you about your \textbf{interaction} with \textbf{technical systems}. The term ‘\textbf{technical systems}’ refers to apps and other \textbf{software applications}, as well as entire \textbf{digital devices} (e.g. mobile phone, computer, TV, car navigation).
        
        Please indicate the \textbf{degree} to which you \textbf{agree/disagree} with the following statements.
    \begin{enumerate}
        \item \label{xhw_study::question::ati_q0}
            I like to occupy myself in greater detail with technical systems.
            \mychoice{completely disagree}
            \mychoice{largely disagree}
            \mychoice{slightly disagree}
            \mychoice{slightly agree}
            \mychoice{largely agree}
            \mychoice{completely agree}
        \item \label{xhw_study::question::ati_q1}
            I like testing the functions of new technical systems.
            \mychoice{completely disagree}
            \mychoice{largely disagree}
            \mychoice{slightly disagree}
            \mychoice{slightly agree}
            \mychoice{largely agree}
            \mychoice{completely agree}
        \item \label{xhw_study::question::ati_q2}
            I predominantly deal with technical systems because I have to.
            \mychoice{completely disagree}
            \mychoice{largely disagree}
            \mychoice{slightly disagree}
            \mychoice{slightly agree}
            \mychoice{largely agree}
            \mychoice{completely agree}
        \item \label{xhw_study::question::ati_q3}
            When I have a new technical system in front of me, I try it out intensively.
            \mychoice{completely disagree}
            \mychoice{largely disagree}
            \mychoice{slightly disagree}
            \mychoice{slightly agree}
            \mychoice{largely agree}
            \mychoice{completely agree}
        \item \label{xhw_study::question::ati_q4}
            I enjoy spending time becoming acquainted with a new technical system.
            \mychoice{completely disagree}
            \mychoice{largely disagree}
            \mychoice{slightly disagree}
            \mychoice{slightly agree}
            \mychoice{largely agree}
            \mychoice{completely agree}
        \item \label{xhw_study::question::ati_q5}
            It is enough for me that a technical system works; I don’t care how or why.
            \mychoice{completely disagree}
            \mychoice{largely disagree}
            \mychoice{slightly disagree}
            \mychoice{slightly agree}
            \mychoice{largely agree}
            \mychoice{completely agree}
        \item \label{xhw_study::question::ati_q6}
            I try to understand how a technical system exactly works.
            \mychoice{completely disagree}
            \mychoice{largely disagree}
            \mychoice{slightly disagree}
            \mychoice{slightly agree}
            \mychoice{largely agree}
            \mychoice{completely agree}
        \item \label{xhw_study::question::ati_q7}
            It is enough for me to know the basic functions of a technical system.
            \mychoice{completely disagree}
            \mychoice{largely disagree}
            \mychoice{slightly disagree}
            \mychoice{slightly agree}
            \mychoice{largely agree}
            \mychoice{completely agree}
        \item \label{xhw_study::question::ati_q8}
            I try to make full use of the capabilities of a technical system.
            \mychoice{completely disagree}
            \mychoice{largely disagree}
            \mychoice{slightly disagree}
            \mychoice{slightly agree}
            \mychoice{largely agree}
            \mychoice{completely agree}
    \end{enumerate}
    \setcounter{qcounter}{\value{enumi}}
\end{enumerate}

\subsection{Microchip Understanding}
\label{xhw_study::subapp::understanding}
\begin{enumerate}
    \setcounter{enumi}{\value{qcounter}}
    \item \label{xhw_study::question::perception}
        What comes to your mind when you think of \textbf{microchips}, also known as "computer chips" and "integrated circuits"? Please take a minute to think about the question and write down everything that comes to your mind.  \textit{[free text]}
    \item \label{xhw_study::question::understand_more}
        Would \textbf{you personally} like to understand more about \textbf{microchips}? 
        \mychoice{Yes, because ...} \textit{[free text]}
        \mychoice{No, because ...} \textit{[free text]}
    \item \label{xhw_study::question::time_willing_before}
        How much time would you \textbf{be willing} to invest \textbf{per newly acquired device} to better understand the \textbf{microchips} it contains?
        \mychoice{less than 1 hour}
        \mychoice{1 to less than 2 hours}
        \mychoice{2 to less than 3 hours}
        \mychoice{3 to less than 4 hours}
        \mychoice{4 or more hours}
    \item \label{xhw_study::question::time_actual}
        How much time do you \textbf{currently} invest \textbf{per newly acquired device} to better understand the \textbf{microchips} it contains?
        \mychoice{less than 1 hour}
        \mychoice{1 to less than 2 hours}
        \mychoice{2 to less than 3 hours}
        \mychoice{3 to less than 4 hours}
        \mychoice{4 or more hours}
    \setcounter{qcounter}{\value{enumi}}
\end{enumerate}

\subsection{A Brief Background on Microchips}
\label{xhw_study::subapp::background}
\begin{enumerate}
    \setcounter{enumi}{\value{qcounter}}
    \item \label{xhw_study::question::background}
        Microchips are tiny objects that store and operate on information in the form of digital data. They are a crucial part of many electronic devices we use every day, like phones, cars, planes, medical implants, and industrial systems. Microchips play a major role in the development of digital technology and make advanced applications like artificial intelligence possible. These chips are highly complex, they are composed of extremely small structures, and are made in various facilities around the world.
    \setcounter{qcounter}{\value{enumi}}
\end{enumerate}

\subsection{Criticality of Use Cases}
\label{xhw_study::subapp::criticality}
\begin{enumerate}
    \setcounter{enumi}{\value{qcounter}}
    \item \label{xhw_study::question::criticality}
        On a scale from from 1---not at all critical to 5---extremely critical, how critical do you \textbf{personally} consider the following \textbf{microchip use cases}. Think about the impact a malfunctioning or failing microchip has \textbf{on you} in each particular use case.
        \begin{enumerate}
            \item \label{xhw_study::question::criticality_airplane}
                You are a passenger in an \textbf{airplane} that contains \textbf{microchips to control its steering}.
                \mychoice{1---not at all critical}
                \mychoice{2---slightly critical}
                \mychoice{3---moderately critical}
                \mychoice{4---very critical}
                \mychoice{5---extremely critical}
            \item \label{xhw_study::question::criticality_car}
                You are driving in a \textbf{car} that contains \textbf{microchips to control its entertainment system}.
                \mychoice{1---not at all critical}
                \mychoice{2---slightly critical}
                \mychoice{3---moderately critical}
                \mychoice{4---very critical}
                \mychoice{5---extremely critical}
            \item \label{xhw_study::question::criticality_smartphone}
                You use a \textbf{smartphone} that contains \textbf{microchips enabling fingerprint unlocking}.
                \mychoice{1---not at all critical}
                \mychoice{2---slightly critical}
                \mychoice{3---moderately critical}
                \mychoice{4---very critical}
                \mychoice{5---extremely critical}
            \item \label{xhw_study::question::criticality_celltower}
                You are making a call through a \textbf{cell tower} that relies on \textbf{microchips for wireless communication}.
                \mychoice{1---not at all critical}
                \mychoice{2---slightly critical}
                \mychoice{3---moderately critical}
                \mychoice{4---very critical}
                \mychoice{5---extremely critical}
            \item \label{xhw_study::question::criticality_pacemaker}
                You have a \textbf{pacemaker} implanted that contains \textbf{microchips to maintain an adequate heart rate}.
                \mychoice{1---not at all critical}
                \mychoice{2---slightly critical}
                \mychoice{3---moderately critical}
                \mychoice{4---very critical}
                \mychoice{5---extremely critical}
        \end{enumerate}
    \setcounter{qcounter}{\value{enumi}}
\end{enumerate}

\subsection{Vignettes}
\label{xhw_study::subapp::vignettes}
\begin{enumerate}
    \setcounter{enumi}{\value{qcounter}}
    \item \label{xhw_study::question::vignete_intro}
        Next, we will show you five different scenarios of devices containing microchips and ask you to answer a few questions for each scenario. 

        Please read the descriptions of each scenario carefully and answer the questions thoughtfully.
    \item \label{xhw_study::question::v0}
        \textbf{Scenario x/5}\\
        Please imagine \textbf{yourself} being in the following situation:\\
        You are a passenger in an \textbf{airplane} that contains microchips to \textbf{control its steering}.\\
        Think about the \textbf{\textcolor[HTML]{B6321C}{safety} implications} of these microchips. Safety means keeping yourself and the system safe from physical harm.\\
        \textcolor{gray}{\infosymbol By hovering over a word marked in red, you can get more information on the respective term.}
        \begin{enumerate}
            \item \label{xhw_study::question::v0_desideratum}
                On a scale from 1---not at all important to 5---extremely \textbf{important}, how important is it to you \textbf{personally} to have a high level of \textbf{\textcolor[HTML]{B6321C}{safety}} for \textbf{microchips controlling the steering of an airplane}?\\
                \textcolor{gray}{\infosymbol When rating the importance of safety in this scenario, you could think about the following questions: Is it relevant to you? Would you care about it?} 
                \mychoice{1---not at all important}
                \mychoice{2---slightly important}
                \mychoice{3---moderately important}
                \mychoice{4---very important}
                \mychoice{5---extremely important}
            \item \label{xhw_study::question::v0_desideratum_open}
                Please briefly explain why you rated the importance of \textbf{\textcolor[HTML]{B6321C}{safety}} to you in this scenario as you did. \textit{[free text]}
            \item \label{xhw_study::question::v0_information}
                On a scale from 1---not at all important to 5---extremely important, how \textbf{important} is it to you \textbf{personally} to receive the following \textbf{information} for assessing the \textbf{\textcolor[HTML]{B6321C}{safety}} of microchips controlling the steering of an airplane?\\
                \textcolor{gray}{\infosymbol When rating the importance of information, you could think about the following questions: Could such information provide any benefit to you? Would they be helpful for you to evaluate the safety?}
            \begin{enumerate}
                \item \label{xhw_study::question::v0_information_who_manucatured}
                    Information about \textbf{\textcolor[HTML]{B6321C}{who designed and manufactured the microchips}}.
                    \mychoice{1---not at all important}
                    \mychoice{2---slightly important}
                    \mychoice{3---moderately important}
                    \mychoice{4---very important}
                    \mychoice{5---extremely important}
                \item \label{xhw_study::question::v0_information_how_interact}
                    Information about \textbf{\textcolor[HTML]{B6321C}{how the microchips interact with the system}}.
                    \mychoice{1---not at all important}
                    \mychoice{2---slightly important}
                    \mychoice{3---moderately important}
                    \mychoice{4---very important}
                    \mychoice{5---extremely important}
                \item \label{xhw_study::question::v0_information_how_approved}
                    Information about \textbf{\textcolor[HTML]{B6321C}{how the microchips have been approved for use}}.
                    \mychoice{1---not at all important}
                    \mychoice{2---slightly important}
                    \mychoice{3---moderately important}
                    \mychoice{4---very important}
                    \mychoice{5---extremely important}
                \item \label{xhw_study::question::v0_information_which_functionality}
                    Information about \textbf{\textcolor[HTML]{B6321C}{which functionality the microchips provide}}.
                    \mychoice{1---not at all important}
                    \mychoice{2---slightly important}
                    \mychoice{3---moderately important}
                    \mychoice{4---very important}
                    \mychoice{5---extremely important}
                \item \label{xhw_study::question::v0_information_how_manucatured}
                    Information about \textbf{\textcolor[HTML]{B6321C}{how the microchips were designed and manufactured}}.
                    \mychoice{1---not at all important}
                    \mychoice{2---slightly important}
                    \mychoice{3---moderately important}
                    \mychoice{4---very important}
                    \mychoice{5---extremely important}
            \end{enumerate}
            \item \label{xhw_study::question::v0_information_open}
                Please briefly explain why you rated the importance of receiving "information about \textbf{\textcolor[HTML]{B6321C}{who designed and manufactured the microchips}}" to you in this scenario as "4---very important".\\
                \textcolor{gray}{\infosymbol By hovering over a word marked in red, you can get more information on the respective term.} \textit{[free text]}
        \end{enumerate}
    \setcounter{qcounter}{\value{enumi}}
\end{enumerate}

\subsection{Microchip Properties}
\label{xhw_study::subapp::properties}
\begin{enumerate}
    \setcounter{enumi}{\value{qcounter}}
    \item \label{xhw_study::question::properties}
        Please assign each \textbf{description} on the left to one of the \textbf{properties} on the right. There is \textbf{one matching description} for each property. If you don't know the assignment, please make a guess. 
        Properties: 
            \mychoice{safety}
            \mychoice{accountability}
            \mychoice{ethical standards} 
            \mychoice{cyber security}
            \mychoice{trustworthiness;}
        descriptions:
            \mychoice{Ensures that microchips do not cause harm to you or the system.}
            \mychoice{Enables figuring out who is responsible in case something goes wrong.}
            \mychoice{Defines practices for responsible treatment of employees and the environment.}
            \mychoice{Makes sure that sensitive information is kept safe from people who are not allowed to see it or change it.}
            \mychoice{Guarantees that a microchip works properly and can also demonstrate this fact.}
    \item \label{xhw_study::question::time_willing_after}
        Now that you have answered the previous questions, how much time would you \textbf{be willing} to invest \textbf{per newly acquired device} to better understand the \textbf{microchips} it contains?
            \mychoice{less than 1 hour}
            \mychoice{1 to less than 2 hours}
            \mychoice{2 to less than 3 hours}
            \mychoice{3 to less than 4 hours}
            \mychoice{4 or more hours}
    \setcounter{qcounter}{\value{enumi}}
\end{enumerate}

\subsection{Demographics}
\label{xhw_study::subapp::demographics}
\begin{enumerate}
    \setcounter{enumi}{\value{qcounter}}
    \item \label{xhw_study::question::gender}
        What is your \textbf{gender}?
            \mychoice{Male}
            \mychoice{Female}
            \mychoice{Non-binary}
            \mychoice{Describe yourself:} \textit{[free text]}
            \mychoice{I prefer not to answer this question}
    \item \label{xhw_study::question::age}
        What is your \textbf{age}?
            \mychoice{18-24}
            \mychoice{25-34}
            \mychoice{35-44}
            \mychoice{45-54}
            \mychoice{55-64}
            \mychoice{65 or older}
            \mychoice{I prefer not to answer this question}
    \item \label{xhw_study::question::education}
        What is your highest level of \textbf{education}?
            \mychoice{High school or equivalent}
            \mychoice{Some college, no degree}
            \mychoice{Associate's degree, occupational}
            \mychoice{Associate's degree, academic}
            \mychoice{Bachelor's degree}
            \mychoice{Master's degree}
            \mychoice{Professional degree}
            \mychoice{Doctoral degree}
            \mychoice{I prefer not to answer this question}
    \item \label{xhw_study::question::experience}
        Do you have practical experience with microchips, \eg, from chip design, manufacturing, testing, deployment, or policies in the semiconductor domain?
            \mychoice{Yes}
            \mychoice{No}
            \mychoice{I prefer not to answer this question}

    \setcounter{qcounter}{\value{enumi}}
\end{enumerate}

\subsection{Feedback}
\label{xhw_study::subapp::feedback}
\begin{enumerate}
    \setcounter{enumi}{\value{qcounter}}
    \item \label{xhw_study::question::feedback}
        Is there anything you would like to tell us about this survey? Please give us your feedback.  \textit{[free text]}
    \item \label{xhw_study::question::outro}
        We thank you for your time spent taking this survey. Your response has been recorded.\\
        Please click the button below to be redirected to Prolific and register your submission.
    \setcounter{qcounter}{\value{enumi}}
\end{enumerate}

\onecolumn
\section{Demographics}
\label{xhw_study::app::demographics}
%See \autoref{xhw_study::tab::demographics} for the demographics of our 250 participants, consisting of gender, age, highest level of education, prior practical experience with microchips, and affinity for technology interaction~\cite{franke2019personal}.

\begin{table}[h]
    \centering
    \footnotesize
    \caption{Demographics of our 250 participants consisting of gender, age, highest level of education, prior practical experience with microchips, and affinity for technology interaction~\cite{franke2019personal}.}
    \label{xhw_study::tab::demographics}
    \Description{The table provides demographic information and is split into two major columns, each comprising three sub-columns. The first row spans both major columns and reads "Demographics (n=250)". Next, in the left major column, the "Gender" and "Age" demographics are provided. "Gender" comprises a header row and five content rows. The header row only spans minor columns 2 and 3 and says "n" and "\%". The next five rows give numbers for "Male", "Female", and "Non-binary" participants, as well as those that chose "Describe yourself" or "No answer". Following the same structure, the "Age" demographics are described using a header row and seven content rows. The header row only spans minor columns 2 and 3 and says "n" and "\%". The next seven rows give numbers for the age brackets from "18-24" up to "65 and older", as well as "No answer". The first demographic in the right major column (starting at minor column 4) is "Education", comprising a header row and nine content rows. The header row only spans minor columns 5 and 6 and says "n" and "\%". The next nine rows give numbers for the different levels of education from "High school or equivalent" to "Doctoral degree", as well as "No answer". Next, the "Prior Experience" demographics are provided, comprising a header row and three content rows. The header row only spans minor columns 5 and 6 and says "n" and "\%". The next three rows give numbers for participants that responded with "Yes", "No", as well as "No answer". Finally, a row describing the participants' "ATI" score spans the entire table across both major columns. It comprises a single row and reads (from left to right) "ATI", "mean", "4.05", "sd", and "0.91".}
    \begin{tabularx}{\linewidth}{Xcc|Xcc}
        \toprule
        \multicolumn{6}{c}{\textbf{Demographics} \textit{(n=250)}}\\
        \midrule
        \textbf{Gender} & \textbf{n} & \textbf{\%} & \textbf{Education} & \textbf{n} & \textbf{\%}\\
        \textit{Male} & 121 & 48.4 & \textit{High school or equivalent} & 35 & 14.0\\ 
        \textit{Female} & 121 & 48.4 & \textit{Some college, no degree} & 54 & 21.6\\
        \textit{Non-binary} & 8 & 3.2 & \textit{Associate's degree, occupational} & 8 & 3.2\\
        \textit{Describe yourself} & 0 & 0.0 & \textit{Associate's degree, academic} & 16 & 6.4\\
        \textit{No answer} & 0 & 0.0 & \textit{Bachelor's degree} & 98 & 39.2\\
        \cmidrule(lr){1-3}
        \textbf{Age} & \textbf{n} & \textbf{\%} & \textit{Master's degree} & 31 & 12.4\\
        \textit{18-24} & 50 & 20.0 & \textit{Professional degree} & 2 & 0.8\\
        \textit{25-34} & 89 & 35.6 & \textit{Doctoral degree} & 4 & 1.6\\
        \textit{35-44} & 57 & 22.8 & \textit{No answer} & 2 & 0.8\\
        \cmidrule(lr){4-6}
        \textit{45-54} & 32 & 12.8 & \textbf{Prior Experience} & \textbf{n} & \textbf{\%}\\
        \textit{55-64} & 15 & 6.0 & \textit{Yes} & 13 & 5.2\\
        \textit{65 or older} & 7 & 2.8 & \textit{No} & 231 & 92.4\\
        \textit{No answer} & 0 & 0.0 & \textit{No answer} & 6 & 2.4\\
        \midrule
        \textbf{ATI} & \textbf{mean} & \multicolumn{1}{c}{4.05} & & \textbf{sd} & 0.91\\
        \bottomrule
    \end{tabularx}
\end{table}

\newpage
\section{Codebooks for Q3 and Q4}
\label{xhw_study::app::codebook}

%For \autoref{xhw_study::question::perception}, see \autoref{xhw_study::tab::codebook_perception}; for \autoref{xhw_study::question::understand_more}, see \autoref{xhw_study::tab::codebook_understand}.

\begin{table*}[htb]
    \centering
    \footnotesize
    \caption{Codebook with absolute and relative code frequencies for 250 responses to \autoref{xhw_study::question::perception}~(\enquote{What comes to your mind when you think of \textbf{microchips}, also known as "computer chips" and "integrated circuits"?}).}
    \label{xhw_study::tab::codebook_perception}
    \Description{The table comes with six columns and 31 rows. The table is split into two major columns with each major column providing a code as well as its absolute and relative frequency per row. The first two rows of each major column are the header rows featuring a column labeled "code" (centered between the two header rows) and two columns jointly labeled "frequency" in the top header row. The "frequency" columns are split in one column labeled "abs." and another one labeled "rel." in the second header row. In all other rows, a code name, the absolute frequency, and the relative frequency of the code are given in each half of the table. The codes are sorted in decreasing order by their frequency.}
    \begin{tabularx}{\linewidth}{L{0.35\linewidth}R{0.04\linewidth}R{0.04\linewidth}|L{0.35\linewidth}R{0.04\linewidth}R{0.04\linewidth}}
        \toprule
        \multirow{2}{*}{\textbf{code}} & \multicolumn{2}{r|}{\textbf{frequency}} & \multirow{2}{*}{\textbf{code}} & \multicolumn{2}{r}{\textbf{frequency}}\\
         & \textit{abs.} & \textit{rel.} & & \textit{abs.} & \textit{rel.}\\
        \midrule
        computer & 104 & 0.42 & circuit & 10 & 0.04 \\
        small size & 84 & 0.34 & manufacturing challenges & 10 & 0.04 \\
        building block that makes things work & 83 & 0.33 & foreign manufacturing & 9 & 0.04 \\
        used across devices & 71 & 0.28 & gaming & 9 & 0.04 \\
        technology & 54 & 0.22 & political aspects & 8 & 0.03 \\
        electronics & 50 & 0.20 & AI & 8 & 0.03 \\
        phone & 47 & 0.19 & circuit board & 8 & 0.03 \\
        technological advancement & 47 & 0.19 & GPU & 8 & 0.03 \\
        data storage & 35 & 0.14 & high complexity & 8 & 0.03 \\
        microchip composition & 35 & 0.14 & tablet & 8 & 0.03 \\
        human implant & 29 & 0.12 & fear & 7 & 0.03 \\
        CPU & 28 & 0.11 & health & 7 & 0.03 \\
        other named devices & 28 & 0.11 & soldering & 7 & 0.03 \\
        animal implant & 27 & 0.11 & vaccines & 6 & 0.02 \\
        data processing & 25 & 0.10 & no idea & 5 & 0.02 \\
        brain similarity & 24 & 0.10 & privacy & 5 & 0.02 \\
        processing power & 22 & 0.09 & binary values & 4 & 0.02 \\
        vehicle & 22 & 0.09 & pop culture reference & 4 & 0.02 \\
        tracking & 18 & 0.07 & security & 4 & 0.02 \\
        supply chain issues & 18 & 0.07 & authentication & 3 & 0.01 \\
        motherboard & 17 & 0.07 & economical dependence & 3 & 0.01 \\
        diverse functionality & 15 & 0.06 & Elon Musk & 3 & 0.01 \\
        memory & 13 & 0.05 & ethical concerns & 3 & 0.01 \\
        companies & 12 & 0.05 & flat & 3 & 0.01 \\
        computer parts & 12 & 0.05 & profitable & 3 & 0.01 \\
        control & 12 & 0.05 & stock market & 3 & 0.01 \\
        communication & 11 & 0.04 & internet & 2 & 0.01 \\
        conspiracy & 11 & 0.04 & toys & 2 & 0.01 \\
        societal impact & 11 & 0.04 &  & \\
        \bottomrule
    \end{tabularx}
\end{table*}

\begin{table*}[htbp]
    \centering
    \footnotesize
    \caption{Codebook with absolute and relative code frequencies for 250 responses to  \autoref{xhw_study::question::understand_more}~(\enquote{Would \textbf{you personally} like to understand more about \textbf{microchips}? Yes/No, because \dots}).}
    \label{xhw_study::tab::codebook_understand}
    \Description{The table comes with six columns and 14 rows. The table is split into two major columns with each major column providing a code as well as its absolute and relative frequency per row. The first two rows of each major column are the header rows featuring a column labeled "code" (centered between the two header rows) and two columns jointly labeled "frequency" in the top header row. The "frequency" columns are split in one column labeled "abs." and another one labeled "rel." in the second header row. In all other rows, a code name, the absolute frequency, and the relative frequency of the code are given in each half of the table. The codes are sorted in decreasing order by their frequency.}
    \begin{tabularx}{\linewidth}{L{0.35\linewidth}R{0.04\linewidth}R{0.04\linewidth}|L{0.35\linewidth}R{0.04\linewidth}R{0.04\linewidth}}
        \toprule
        \multirow{2}{*}{\textbf{code}} & \multicolumn{2}{r|}{\textbf{frequency}} & \multirow{2}{*}{\textbf{code}} & \multicolumn{2}{r}{\textbf{frequency}}\\
         & \textit{abs.} & \textit{rel.} & & \textit{abs.} & \textit{rel.}\\
        \midrule
        gain knowledge & 96 & 0.38 & application areas & 12 & 0.05\\
        understand functionality & 46 & 0.18 & impact on society & 10 & 0.04\\
        omnipresent in daily life & 32 & 0.13 & understand manufacturing & 10 & 0.04\\
        no interest & 28 & 0.11 & professional needs & 9 & 0.04\\
        incomplete knowledge & 24 & 0.10 & satisfied & 9 & 0.04\\
        scientific progress & 24 & 0.10 & fear & 8 & 0.03\\
        keep up with progress & 20 & 0.08 & informed decision making & 7 & 0.03\\
        operation before knowledge & 18 & 0.07 & risk assessment & 7 & 0.03\\
        no need & 16 & 0.06 & diagnose issues & 6 & 0.02\\
        using technology & 16 & 0.06 & improve productivity & 4 & 0.02\\
        too complicated & 15 & 0.06 & improve quality of life & 4 & 0.02\\
        importance for future & 14 & 0.06 & explain to others & 2 & 0.01\\
        \bottomrule
    \end{tabularx}
\end{table*}

\newpage
\section{Detailed Results of Multilevel Regression Analysis}
\label{xhw_study::app::regression_results}

\begin{table*}[htbp]
    \centering
    \footnotesize
    \caption{Multilevel regression analysis, including interactions, based on participants' ratings of the importance of receiving different types of information to evaluate desideratum in a given setting, on a scale from \textit{1---not at all important} to \textit{5---extremely important} (see \autoref{xhw_study::question::v0_information} for an example question presented to participants).}
    \label{xhw_study::tab::detailed_regression_information}
    \Description{The table is organized in six columns. In the first column on the left, the predictors for multilevel regression analysis are given. Each of the other five major columns provides details on a type of information about microchips, starting with "which functionality" on the left and ending with "how manufactured" on the right. Each row of the five major columns gives the Estimate ("Est.") for one predictor. Here, "*" means p<0.05, "**" means p<0.01, and "***" means p<0.001, and respective entries are highlighted in boldface. Next, the intercept of regression analysis is given for each kind of information. Next, a sub-header spanning the entire table titled "interactions (baseline=car x ethical standards)" opens the section with the analysis results for the interactions between settings and desiderata. The estimates for these interactions are given in the next 24 rows. In the next section, the first row is titled "desire to understand more about microchips" and lists respective estimates. The second and final row of this section is titled "ATI score" and lists respective estimates. Finally, in the last section of the table, the first row lists results for "marginal R^2" and the second row "conditional R^2".}
    \begin{tabularx}{\linewidth}{XL{1.5cm}L{1.5cm}L{1.5cm}L{1.5cm}L{1.5cm}}
        \toprule
         & \textbf{which func-} & \textbf{how} & \textbf{how} & \textbf{how manu-} & \textbf{who manu-}\\
         & \textbf{tionality} & \textbf{interacts} & \textbf{approved} & \textbf{factured} & \textbf{factured}\\
         \textit{Predictors} & \textit{Est.} & \textit{Est.} & \textit{Est.} &  \textit{Est.} & \textit{Est.} \\
         \midrule
        intercept: car (setting) $\times$ ethical standards (desideratum) & 2.87*** & 2.43***  & 2.82***  & 2.71***  & 2.66*** \\
        \midrule
        \textit{interactions (baseline=car $\times$ ethical standards)}\\
        car $\times$ accountability & -0.16 & -0.01 &  -0.21 & -0.37  & 0.00 \\
        car $\times$ safety & -0.25 & 0.09 & -0.10 &  -0.33 & -0.08 \\
        car $\times$ trustworthiness & -0.03 & 0.04 & \textbf{-0.56*} & \textbf{-0.65**} & \textbf{-0.53*}\\
        car $\times$ cyber security & 0.12 &  \textbf{0.65**} & 0.36 & -0.03 & 0.11 \\
        smartphone $\times$ ethical standards & 0.02  & 0.25 & 0.37 & 0.29 & \textbf{0.60*}\\
        smartphone $\times$ accountability & 0.22 & 0.07 & 0.06  & 0.17 &  -0.06 \\
        smartphone $\times$ safety & 0.18  & 0.05 & -0.04 & -0.07 & -0.53\\
        smartphone $\times$ trustworthiness & 0.44  & 0.13  & 0.39  & 0.08 &  -0.05 \\
        smartphone $\times$ security & 0.32 & -0.07 & -0.48 & -0.40 & -0.52 \\
        cell tower $\times$ ethical standards & -0.11 & 0.31 & 0.28 & 0.39 & \textbf{0.63*} \\
        cell tower $\times$ accountability & 0.24 & -0.01 & -0.39 & -0.62 &  \textbf{-0.93*}\\
        cell tower $\times$ safety & 0.47 &  -0.10 & 0.27 & -0.18 & -0.64 \\
        cell tower $\times$ trustworthiness & 0.28 & -0.02 & 0.28 & 0.12 &  -0.16\\
        cell tower $\times$ security & 0.33 & -0.35 & 0.15 & -0.19  & -0.21 \\
        pacemaker $\times$ ethical standards & -0.11 & 0.31 & 0.28 & 0.39 & \textbf{0.63*} \\
        pacemaker $\times$ accountability & 0.42  & 0.67 & 0.60 & 0.19 & 0.25 \\
        pacemaker $\times$ safety & \textbf{0.88*} & 0.49 & 0.41 & 0.53  & 0.16 \\
        pacemaker $\times$ trustworthiness & 0.60 & 0.67 & \textbf{1.23**} & \textbf{1.23**} & \textbf{1.12**} \\
        pacemaker $\times$ security & 0.47 &  -0.04 & 0.10 & 0.18  & 0.14 \\
        airplane $\times$ ethical standards & 0.20 & \textbf{0.53*} & 0.48 & \textbf{0.81**} & \textbf{0.78**} \\
        airplane $\times$ accountability & 0.59  & 0.67 & 0.61 & 0.19 & 0.25 \\
        airplane $\times$ safety & 0.43 & 0.01 & 0.52 & -0.38  & -0.34 \\
        airplane $\times$ trustworthiness & 0.46 & 0.31 & \textbf{0.99**} & 0.34 & 0.30 \\
        airplane $\times$ security & 0.31 &  -0.07 & -0.22 & -0.61  & -0.48 \\
        \midrule
        desire to understand more about microchips & \textbf{0.54***} & \textbf{0.63***} & \textbf{0.40*} & \textbf{0.49**} & \textbf{0.41*}\\
        ATI score & \textbf{0.26***} & \textbf{0.27***} & \textbf{0.26**} & \textbf{0.28***} & \textbf{0.26**}\\
        \midrule
        marginal $R^2$    & 0.177 & 0.198 & 0.180 & 0.189 & 0.155\\
        conditional $R^2$ & 0.465 & 0.478 & 0.522 & 0.571 & 0.536\\    
        \bottomrule
     \end{tabularx}
     \textit{* p < 0.05, ** p < 0.01, *** p < 0.001}
\end{table*}
%TC:endignore

\end{document}